\documentclass[10pt,journal,compsoc]{IEEEtran}
\usepackage{ifpdf}
\ifCLASSOPTIONcompsoc
  \usepackage[nocompress]{cite}
\else
  \usepackage{cite}
\fi
\usepackage{amsmath}
\usepackage{amsopn}
\usepackage{subcaption}
\usepackage{algorithmicx}
\usepackage{algorithm} 
\usepackage{algpseudocode} 
\usepackage{array}
\usepackage{url}
\usepackage{booktabs}
\usepackage{multirow}
\usepackage{float}
\usepackage[table,xcdraw]{xcolor}
\usepackage{gensymb}
\usepackage{textcomp}
\usepackage{graphicx}
\usepackage[outdir=./]{epstopdf}
\graphicspath{ {./images/} }
\usepackage{amssymb}
\usepackage{caption}
\usepackage{calc}
\usepackage{comment}
\usepackage[table,xcdraw]{xcolor}
\usepackage{cite}
\def\BibTeX{{\rm B\kern-.05em{\sc i\kern-.025em b}\kern-.08em
    T\kern-.1667em\lower.7ex\hbox{E}\kern-.125emX}}
 
\begin{document}

\title{DeeBBAA: A benchmark \underline{Dee}p \underline{B}lack \underline{B}ox \underline{A}dversarial \underline{A}ttack against Cyber-Physical Power Systems}

\author{Arnab~Bhattacharjee,~Tapan~K.~Saha,~\IEEEmembership{Fellow,~IEEE}, ~Ashu~Verma,~\IEEEmembership{Senior Member,~IEEE},~Sukumar~Mishra,~\IEEEmembership{Senior Member,~IEEE}

\thanks{Arnab Bhattacharjee is enrolled as a Ph.D. student in Electrical Engineering at the University of Queensland-IIT Delhi Academy of Research. email:  Arnab.Bhattacharjee@uqidar.iitd.ac.in}%
\thanks{Tapan K Saha is at the School of Information Technology and Electrical Engineering in the University of Queensland, Brisbane, Australia. email: saha@itee.uq.edu.au}%
\thanks{Ashu Verma is associated with the Department of Energy Science and Engineering, Indian Institute of Technology, Delhi, India, 110016. email: averma@dese.iitd.ac.in}%
\thanks{Sukumar Mishra is associated with the Electrical Engineering Department, Indian Institute of Technology, Delhi, India, 110016. email: sukumar@ee.iitd.ac.in}}%


\IEEEtitleabstractindextext{%
\begin{abstract}
An increased energy demand, and environmental pressure to accommodate higher levels of renewable energy and flexible loads like electric vehicles have led to numerous smart transformations in the modern power systems. These transformations make the cyber-physical power system highly susceptible to cyber-adversaries targeting its numerous operations. In this work, a novel black box adversarial attack strategy is proposed targeting the AC state estimation operation of an unknown power system using historical data. Specifically, false data is injected into the measurements obtained from a small subset of the power system components which leads to significant deviations in the state estimates. Experiments carried out on the IEEE 39 bus and 118 bus test systems make it evident that the proposed strategy, called DeeBBAA, can  evade numerous conventional and state-of-the-art attack detection mechanisms with very high probability.

\end{abstract}

}

\maketitle
\IEEEdisplaynontitleabstractindextext
\IEEEpeerreviewmaketitle

\IEEEraisesectionheading{\section{Introduction}\label{sec:intro}}

\IEEEPARstart{G}{rowing} energy needs of the 21st century, supplemented with increasing environmental pressure to replace conventional methods of energy generation with highly intermittent and unreliable renewable energy sources, have led to significant transformations in conventional power systems. Over the last few years, the confluence of information and automation technologies has improved the reliability, resilience, and operational efficiency of these modern conduits of energy, rendering them a dynamic cyber-physical nature \cite{9272724}. In order to monitor and operate the power grid efficiently and prepare it against potential contingencies, power system operators regularly perform a critical task known as the Power System State Estimation (PSSE), which involves filtering and processing measurements from grid sensors in a centralized control center. The measurements are collected by a Supervisory Control and Data Acquisition (SCADA) system and communicated to the control center via Remote Terminal Units (RTUs). The real-time operating status of the grid obtained from the outcome of the PSSE is then used to improve situational awareness, make economic decisions about energy dispatch, and take contingency actions against threats that endanger the reliability of the grid \cite{8403288}.

In a smart cyber-physical power system where remote transmission of crucial information is carried out, it is imperative to ensure the security of the communication channels to protect the system against intrusion and subsequent degradation of operational integrity. In the face of growing threats from cyberattacks, like the one that led to the disruption of power distribution in Ukraine in 2015 \cite{Ukraine_attack}, conventional defenses like firewalls and air gaps are deemed inadequate. Under these circumstances, envisioning and anticipating the myriad ways remote attackers can target these cyber-physical systems has become a need of the hour \cite{NAP24836}. 

The objective of this study is to analyze the vulnerability of a smart cyber-physical power system, particularly the state estimation operation, against a specific type of cyber-attack called the Stealthy False Data Injection Attack (SFDIA) that compromises data integrity while being able to evade Bad Data Detection (BDD) by the control center  and can cause load shedding, economic loss, and even blackouts \cite{7438916, 8403288}. A novel methodology leveraging transfer-based black box adversarial attack generation is presented for stealthy false data injection where the adversary is assumed to have no knowledge of the targeted power system other than access to incomplete historical data pertaining to a subset of power system components. Further analysis shows that the attacks generated by the proposed approach can bypass a variety of state-of-the-art Bad Data Detection algorithms, including data-driven methods and physics-inspired statistical consistency checking algorithms, with very high probability.

\subsection{Related Works}
To carry out SFDIA, an attacker needs to compromise real-time power measurements by intruding into the communication network between the remote terminal units and the control center. While numerous works addressing the design of SFDIAs exists, most of these primarily study a simplified model of the power system based on a linear DC power flow (\cite{10.1145/1952982.1952995, 5754636, 9146569}). In a more practical AC power flow setting, where the power system measurements are a non-linear function of the power system states, the design of SFDIAs becomes difficult due to the spurious nature of the non-linear AC power flow model \cite{8403288, 6805238, 8360557}. Indeed, SFDIAs built using the linearized DC power flow model (dc-SFDIA) can be easily detected by an AC power flow-based BDD (ac-BDD) \cite{WANG20131344}. Power systems are thus more vulnerable to ac-SFDIAs that can bypass ac-BDD and hence are more relevant to this study. 

The naive way to design ac-SFDIA requires the attacker to have complete knowledge of the target power network, including its topology and line parameters, and access to real-time state estimates, which are both highly impractical assumptions. Slightly relaxing the omnipotence assumption on the attacker, early works like \cite{6805238, 7401110, 8260948, 9070219, 8425789, 9239328, 7366616} showed that strong ac-SFDIA can be carried out even if the attacker has access to only the model information pertaining to the entire or a localized region of the targeted power network. However, given the highly classified nature of such power network model information, it still requires the attacker to be an extremely powerful and well-connected entity.

Completely doing away with the omnipotence assumptions, recent research works have proposed model-free methods that use historical power system data to generate ac-SFDIA against an unknown power system model. Using historical data to estimate network parameters, targeted blind FDIAs were generated in \cite{9518372}. \cite{7934033} formulated the ac-SFDIA generation task as a constrained optimization task on the measurement vector and used a geometric approach to derive sufficient conditions for evading BDD. A similar approach using principal component analysis was proposed in \cite{7001709}. Generative modeling is another widely used tool for data driven SFDIA generation. Wasserstein GAN \cite{9853635} and a self-attention-based GAN \cite{9506908} frameworks were proposed in the literature for modeling ac-SFDIA against an unknown power system. A major limitation of the aforementioned works is that their evasive properties are tested against a very limited set of Bad Data Detection algorithms, primarily the conventional Largest Normalized Residue Test and the Chi-Squared Test algorithms. Numerous powerful statistical methods and data-driven algorithms for SFDIA detection have since been proposed in the literature, and whether these ac-SFDIA design strategies are capable of bypassing them is not known.

Over the years, learning-based algorithms have found applications in various operations of the cyber-physical power system, primarily catering to state estimation, forecasting, and control operations\cite{8403288}. Adversarial attacks threaten the operational integrity of such algorithms\cite{9622117}. Depending on whether the attacker has complete knowledge of the target model, adversarial attacks can be either white-box or black-box in nature\cite{https://doi.org/10.48550/arxiv.1810.00069, Wiyatno2019AdversarialEI}. Black box attacks are more practical, albeit more difficult to design, as the target model is unknown. Designing black-box attacks against cyber-physical systems like a power system comes with added difficulties because even though adversarial attacks are demonstrated to be extremely effective against data-driven models, they can easily violate the physics of the power network, thus risking easy detection by BDD safeguards that carry out statistical and physical consistency checks of the system measurements data\cite{9622117, 9609659, 9762566}.

The objective of this work is to design a black box adversarial attack strategy targeting the state estimation operation of an unknown power system, using incomplete historical data pertaining to a small subset of the power system components in such a way that they can evade both data-driven and physics-inspired conventional and statistical Bad Data Detectors. Two existing works in literature (\cite{9737024, 9695995}) are closely related to the proposed work. In \cite{9737024}, a sophisticated GAN framework was proposed to generate black box adversarial perturbations against data-driven control strategies in a power system, particularly targeting transient stability control. The adversarial perturbations were designed specifically to misdirect data-driven control algorithms while bypassing conventional BDD. A major limitation of the algorithm is that it needs to query the target model and use the corresponding outputs during the offline training stage of the GAN, which is highly impractical primarily because querying an online control or estimation algorithm in a power system without raising suspicions is almost impossible and also because no restrictions are imposed on the number of queries required to be made. Unlike this, the proposed method needs limited historical data from the targeted unknown power system only once during training in order to develop a substitute model partially mimicking the unknown state estimation operation, instead of continuous querying. In \cite{9695995} a joint adversarial attack and stealthy false data injection attack was proposed in an attempt to bypass both conventional and data-driven BDD. A DC power flow model was used for SFDIA generation, and the attacker was assumed to have complete knowledge of the power system model and the data-driven BDD, which was formulated using a Multi-layer Perceptron classifier. The authors also demonstrated the lack of transferability of their proposed attack strategy against unknown SFDIA detectors in the black box setting, unlike in the proposed method which is a completely black box SFDIA generator that is demonstrated to be able to evade multiple BDD safeguards.

\subsection{Key Contributions}
The key contributions of this work are enlisted below:
\begin{enumerate}
    \item DeeBBAA is a benchmark black box adversarial attack strategy that directly targets an unknown non-linear AC-PSSE module of a cyber -physical power system. Using a substitute neural regression network, it mimics the physics behind the AC-PSSE operation using incomplete historical data following which adversarial examples generated against the substitute model are used as stealthy FDI attack vectors. No querying interaction is necessary between the attacker and the target model during training or inference stages.
    \item A generic strategy is presented for developing adversarial examples targeting a regression network following which a novel convex reformulation is proposed for fast attack generation. The spread and magnitude of the attack can be adjusted using user-tunable parameters.
    \item Unlike existing works where adversarial attacks are generated against specific data-driven target models, the attacks carried out using DeeBBAA can evade a wide range of SFDIA detection algorithms including physics-inspired detectors like the conventional residue-based BDD algorithms like LNRT and $\chi^2$ test, and statistical consistency-evaluating methods as well as supervised and unsupervised learning-based attack detectors.
    \item The attacker neither requires any privileged power system information like network topology or line parameters, nor do they have the requirement to know the SFDIA detection algorithms put in place by the operator for safeguarding the power network against data-manipulation. Only one time incomplete historical power system data pertaining to a small subset of the power system components is required to design the attacks.
    \item The adversary can modulate the dimension of the attack at will, i.e., it can target any random subset of power system components. Two broad types of target/attack regions are considered in this work, providing greater flexibility to the adversary in a way that they can choose to attack components that are relatively less secure than others.
    \item The attack model is versatile, i.e, can be used for targeting any system from which one-time partial and limited historical data can be accessed by an adversary during reconnaissance and is transferable, scalable and computationally efficient.

\end{enumerate}

 
\subsection{Organization}
The rest of the paper is organised as follows: Section \ref{sec:Background} provides a brief introduction to power system fundamentals, AC state estimation, conventional bad data detection algorithms and adversarial attacks. Section \ref{sec:Proposed method} details the proposed attack model. In section \ref{sec:case}, simulation and experimental details and results corresponding to the implementation of DeeBBAA  are presented. The paper is concluded in Section \ref{sec:conclusion}.
\subsection{Notations}
$\mathbb{R}$ and $\mathbb{C}$ denote the sets of real and complex numbers. $\mathbb{S}^n$ and $\mathbb{H}^n$ represent the sets of $n\times n$ real symmetric and complex Hermitian matrices respectively. $[n]$ denotes the index set ${1,2,...,n}$. $\#(.)$ gives the cardinality of a set. Vectors are written in lowercase bold letters, matrices are represented using bold uppercase letters. $\pmb{0}_n$, $\pmb{1}_n$, $\pmb{0}_{m\times n}$, $\pmb{I}_{n\times n}$ denote the $1\times n$ zero vector, the $1\times n$ one vector, the $m\times n$ zero matrix and the $n\times n$ identity matrix respectively. $W_{ij}$ and $v_j$ represents the $(i,j)$-the entry of matrix $\pmb{W}$ and the $j$-th element of vector $\pmb{v}$ respectively. $\pmb{W}\succeq 0$ implies that $\pmb{W}$ is Hermitian and positive semidefinite(PSD). $(.)^T$ and $(.)^*$ represent the transpose and conjugate transpose operators. $\Re (.)$, $\Im(.)$ and $Tr(.)$ determine the real part, imaginary part and trace of a scalar/matrix respectively. $\tilde{j}=\sqrt{-1}$. The Hadamard product is represented using $\odot$. $\angle{x}$ and $|x|$ represent the angle and magnitude of a complex scalar. $||.||_p$ represents the p-norm.  

\section{Background}\label{sec:Background}
\subsection{Power System Modelling}\label{subsec:PS_analysis}
\begin{figure}[!ht]
    \centering
    \includegraphics[ keepaspectratio,width=0.7\columnwidth]{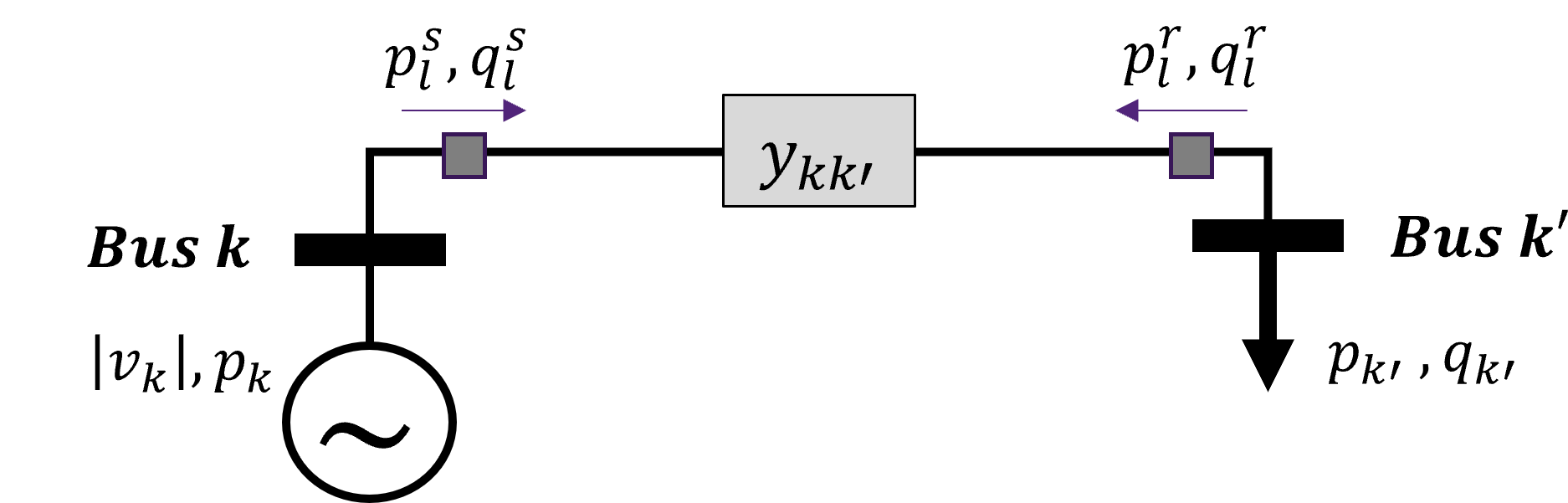}
    \caption{Depiction of a simple 2 bus system with $N = 2$ and $M = 1$. A load with real and reactive power, $p_{k'}$ and $q_{k'}$ is connected to bus $k'$ and a generator, with measurements of real nodal power injection, $p_k$ and voltage magnitude $|v_k|$ to bus $k$. The branch power flows over line $l$ are also shown.}
    \label{fig:PS_funda}
\end{figure}

The electric power grid can be modelled as a graph $\mathcal{G}:= \{\mathcal{N}, \mathcal{L}\}$, where $\mathcal{N}:= [N]$ and $\mathcal{L}:= [M]$ represents its set of buses and branches respectively. Each branch $l \in \mathcal{L}$, where $l$ connects nodes $s$ to $t$, is characterized by its admittance, $y_{st}$ that represents the ease with which electrical current can flow through the branch. The branch admittance matrix $\pmb{Y} \in \mathbb{C}^{N\times N}$ can be understood as a weighted laplacian matrix of $\mathcal{G}$ whose weights are given by the admittances in each branch. The from and to branch admittance matrices $\pmb{Y_s} \in \mathbb{C}^{M\times N}$ and $\pmb{Y_r} \in \mathbb{C}^{M\times N}$ represent weighted branch to node incidence matrices. The power grid topology is implicit in $\pmb{Y}, \pmb{Y_s}$ and $\pmb{Y_r}$. Referring to the 2-bus system in Fig. \ref{fig:PS_funda}, $\pmb{Y}, \pmb{Y_s}$ and $\pmb{Y_r}$ can be written as shown below:

\begin{align}
    \pmb{Y} =   \begin{bmatrix}
                y_k+y_{kk'} & -y_{kk'} \\
                -y_{kk'} & y_{k'}+y_{kk'} 
                \end{bmatrix}
\end{align}

\begin{align}
    \pmb{Y_s} =   \begin{bmatrix}
                y_k+y_{kk'} & -y_{kk'} 
                \end{bmatrix} , \pmb{Y_r} =   \begin{bmatrix}
                -y_{kk'} & y_{k'}+y_{kk'} 
                \end{bmatrix}
\end{align}
where $y_k$ represents the admittance-to-ground at bus $k$. For more details on the construction of the $\pmb{Y}, \pmb{Y_s}$ and $\pmb{Y_r}$, refer \cite{Stevenson}. 

The state of a power system, denoted by $\pmb{x}$, consists of the bus voltage vector $\pmb{x} = \pmb{v} = [v_1, \cdots, v_N]^T \in \mathbb{C}^N$, where $v_j \in \mathbb{C}$ is the complex voltage at bus $j \in \mathcal{N}$ with magnitude $|v_j|$ and phase $\theta_j = \angle{v_j}$. The nodal current injection vector $\pmb{i}=\pmb{Y}\pmb{v}$. The sending and receiving end branch currents are given by $\pmb{i}_s=\pmb{Y_s}\pmb{v}$ and $\pmb{i}_r=\pmb{Y_r}\pmb{v}$, respectively. Let $\{\pmb{a}_1,\cdots,\pmb{a}_N\}$ and $\{\pmb{b}_1,\cdots,\pmb{b}_M\}$ be the sets of canonical vectors in $\mathbb{R}^N$ and $\mathbb{R}^M$, respectively. Power system measurements include voltage magnitude and real and reactive power injection measurements at each bus and the real and reactive sending and receiving end branch power flow measurements at each bus. The measurements can be represented as non-linear functions of the system state using AC power flow equations as shown below:
\begin{enumerate}
    \item \textit{Voltage Magnitude}: The voltage magnitude at bus $k$ is given by $|v_k| = \sqrt{\Re(v_k)^2+\Im(v_k)^2}$
    \item \textit{Bus Power Injection}: The complex power injection at bus $k$ consisting of real and reactive powers, $p_k+\tilde{j}q_k$, where $\pmb{A}_k= \pmb{a}_k\pmb{a}_k^T$, is given by:
    \begin{equation}
        \begin{aligned}
        p_k &= \Re({i_k^*v_k}) = Tr(\frac{1}{2} (\pmb{Y}^*\pmb{A}_k + \pmb{A}_k\pmb{Y})\pmb{v}\pmb{v}^*) \\[-1pt]
        q_k &= \Im({i_k^*v_k}) = Tr(\frac{1}{2\tilde{j}} (\pmb{Y}^*\pmb{A}_k - \pmb{A}_k\pmb{Y})\pmb{v}\pmb{v}^*)\\[-1pt]
        \end{aligned}
        \label{eq:acpf}
    \end{equation}
    \item \textit{Branch Power Flows}: The sending and receiving end power flows in a branch $l \in \mathcal{L}$ connecting nodes $k$ to $k'$ are given by:
    
    \begin{equation}
    \begin{aligned}
        p_l^s &= \Re([\pmb{i_s}]_l^*v_k) = Tr(\frac{1}{2} (\pmb{Y_s}^*\pmb{b}_l\pmb{a}_k^T + \pmb{a}_k\pmb{b}_l^T\pmb{Y_s})\pmb{v}\pmb{v}^*)\\[-1pt]
        p_l^r &= \Re([\pmb{i_s}]_l^*v_{k'}) = Tr(\frac{1}{2} (\pmb{Y_s}^*\pmb{b}_l\pmb{a}_{k'}^T + \pmb{a}_{k'}\pmb{b}_l^T\pmb{Y_s})\pmb{v}\pmb{v}^*)\\[-1pt]
        q_l^s &= \Im([\pmb{i_s}]_l^*v_k) = Tr(\frac{1}{2\tilde{j}} (\pmb{Y_s}^*\pmb{b}_l\pmb{a}_k^T - \pmb{a}_k\pmb{b}_l^T\pmb{Y_s})\pmb{v}\pmb{v}^*)\\[-1pt]
        q_l^r &= \Im([\pmb{i_s}]_l^*v_{k'}) = Tr(\frac{1}{2\tilde{j}} (\pmb{Y_s}^*\pmb{b}_l\pmb{a}_{k'}^T - \pmb{a}_{k'}\pmb{b}_l^T\pmb{Y_s})\pmb{v}\pmb{v}^*)\\[-1pt]
    \end{aligned}
    \label{eq:aclf}
\end{equation}
\end{enumerate}
Let $\pmb{z} = [\{|v_k|, p_k, q_k\}_{\forall k \in \mathcal{N}}, \{p_l^s, q_l^s, p_l^r, q_l^r\}_{\forall l \in \mathcal{L}}]$ $\in \mathbb{R}^{N_m}$, where $N_m = 3\times N + 4\times M$, represent the vector consisting of all power system measurements and consider that the complete set of non-linear AC power flow equations mapping the states to the measurements as shown above can be compactly denoted by $\pmb{h}: \mathbb{C}^{N} \to \mathbb{R}^{N_m}$, then the forward relation between power system measurements and states can be written as shown below:
\begin{equation}
    \begin{aligned}
         \pmb{z} = \pmb{h(x)} + \zeta\\
    \end{aligned}
    \label{eq:z_h_x}
\end{equation}

where $\zeta$ is the measurement noise. In the subsequent sections, an alternative representation of the state vector as shown below will be used instead of the complex representation:
\begin{align*}
    \pmb{x} = [|v_1|,\cdots,|v_N|,\theta_1,\cdots, \theta_N] \in \mathbb{R}^{2N}
\end{align*}

\subsection{AC State Estimation}\label{subsec: acse}
Voltage and power measurements from buses and lines are collected using sensors and sent to the SCADA system through Remote Terminal Units (RTUs). These are then sent to the control center over communication channels where state estimation is executed. Given the noisy measurements, $\pmb{z}$, an iterative weighted least squares problem is solved by the non-linear AC-PSSE module to obtain accurate estimates of the states as shown below:
\begin{equation}
    \begin{aligned}
        \hat{\pmb{x}} &= \mathop{\textrm{min}}_{\pmb{x}} [(\pmb{z}-\pmb{h}(\pmb{x}))^T \mathnormal{\pmb{R}}^{-1} (\pmb{z}-\pmb{h}(\pmb{x}))]  
    \end{aligned}
    \label{eq:acse}
\end{equation}

where $\pmb{R}$ is the error covariance matrix of the measurement vector and $\hat{\pmb{x}}$ represents the estimated state vector.

\subsection{Conventional Bad Data Detection Algorithms}\label{subsec:def_bdd}
\begin{figure}[!ht]
    \centering
    \includegraphics[ keepaspectratio,width=0.8\columnwidth]{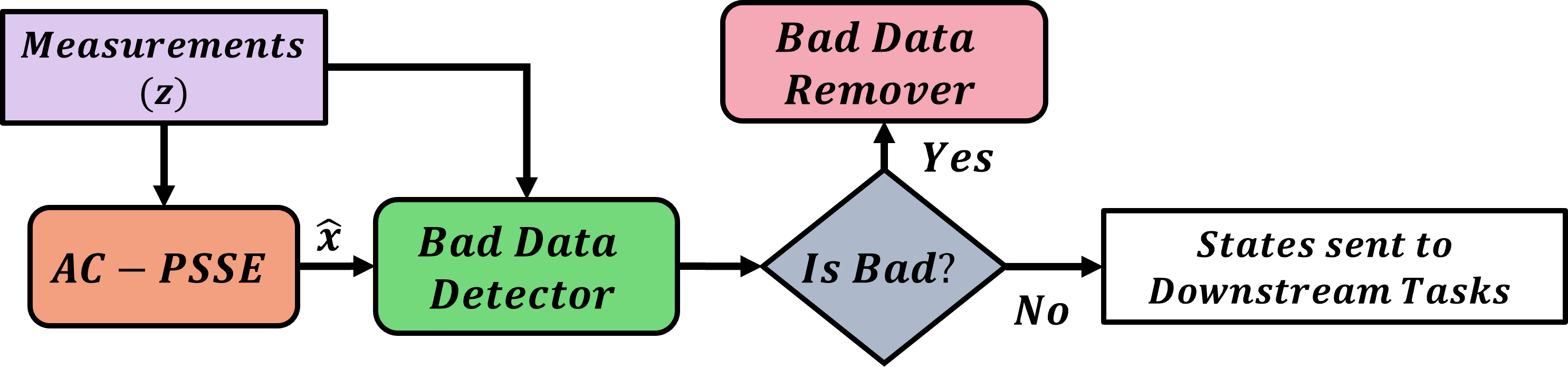}
    \caption{Flow diagram of Conventional Bad Data Detection in Power systems.}
    \label{fig:BDD}
\end{figure}

In order to detect the presence of deliberately induced false data in the measurement vectors, power system operators conventionally employ Bad Data Detection(BDD) algorithms to isolate bad data and prevent false state estimates from being used for downstream tasks as shown in Fig. \ref{fig:BDD}. Conventional BDD strategies include the Largest Normalized Residue test (LNRT) and the $\chi^2$ test. The procedure involves computing the difference between the measurement values collected from the power network, represented by $\pmb{z}$ and those computed from the estimated states, $\hat{\pmb{x}}$ given by $\pmb{h}(\hat{\pmb{x}})$. After obtaining the measurement residue vector, $\pmb{r}$, it is normalized and a hypothesis testing is carried out on its norm, $||\pmb{r}^{norm}||$. Depending on whether the $L_2$ norm or the $L_\infty$ norm is used for the hypothesis testing, it is called $\chi^2$ test or LNRT.

A conventional BDD algorithm can be represented using the following equations:

\begin{equation}
    \begin{aligned}
    &\pmb{r} = |\pmb{z} - h(\hat{\pmb{x}})| \\[-1pt]
    &\mathnormal{\pmb{K}} = \mathnormal{\pmb{H}}^T\mathnormal{\pmb{R}}^{-1}\mathnormal{\pmb{H}}\\[-1pt]
    &\mathnormal{\pmb{B}} = \mathnormal{\pmb{I}} - \mathnormal{\pmb{H}}(\mathnormal{\pmb{K}}\mathnormal{\pmb{H}}^T\mathnormal{\pmb{R}}^{-1})\\[-1pt]
    &r_i^{norm} = \frac{|z_i-h(\hat{x})_i|}{\mathnormal{R}_{ii}\mathnormal{B}_{ii}}
    \end{aligned}
    \label{eq:bdd1}
\end{equation}

The null and alternative hypotheses are given by $\mathcal{H}_0$: Attack has not taken place and $\mathcal{H}_1$: System has been attacked. The $\chi^2$ test is given by:

\begin{equation}
    \begin{aligned}
	 &\hspace*{14mm} \mathcal{H}_1 \\[-1pt]
	 & ||\pmb{r}^{norm}||_{2} \gtrless \tau_{2} \\[-1pt]
	 &\hspace{14mm} \mathcal{H}_0 \\[0pt]
    \end{aligned}
    \label{eq:chi_squared}
\end{equation}

The LNRT is given by:

\begin{equation}
    \begin{aligned}
	 &\hspace*{15mm} \mathcal{H}_1 \\[-1pt]
	 & ||\pmb{r}^{norm}||_{\infty} \gtrless \tau_{\infty} \\[-1pt]
	 &\hspace{15mm} \mathcal{H}_0 \\[-1pt]
    \end{aligned}
    \label{eq:LNRT}
\end{equation}

$\tau_{2}$ and $\tau_{\infty}$ are pre-specified thresholds that are obtained by fixing a specific false alarm rate.

\subsection{Stealthy False Data Injection Attacks}\label{subsec:def_fdia}
In an FDIA, the attacker adds a malicious vector to the acquired measurements to force the AC-PSSE to converge to a different state value than expected. 
\begin{align*}
    \pmb{z}_a = \pmb{z} + \pmb{a} = \pmb{h}(\hat{\pmb{x}}+\pmb{c}) + \zeta
\end{align*}
 $\pmb{a}$ is the attack vector added to the original measurements $\pmb{z}$ leading to the introduction of an unsolicited deviation in the estimated states represented by $\pmb{c}$. A stealthy FDIA vector can bypass the conventional BDD with very high probability. 
With an AC power system model, a sufficient condition on the attack vector $\pmb{a}$ for bypassing the conventional residue based BDD is given by \cite{6275516}:
\begin{align}
    \pmb{a} = \pmb{h}(\hat{\pmb{x}} + \pmb{c})-\pmb{h}(\hat{\pmb{x}}) \label{eq:SFDIA}
\end{align}
since if $\pmb{r_i}$ and $\pmb{r_f}$ denote the measurement residue vectors before and after SFDIA, then:
\begin{align}
    \pmb{r_f} = \pmb{z}_a - \pmb{h}(\hat{\pmb{x}}+\pmb{c}) = \pmb{z} + \pmb{a} - \pmb{h}(\hat{\pmb{x}}+\pmb{c}) = \pmb{z} - \pmb{h}(\hat{\pmb{x}}) = \pmb{r_i}
 \label{eq:same_res}
\end{align}
implying that the conventional residue based BDD won't be able to distinguish an attacked sample from a benign one. However in order to compute $\pmb{a}$ using Eq. \ref{eq:SFDIA}, the attacker needs to have complete knowledge of the power system without which $\pmb{h}(.)$ cannot be computed. Also, the real-time state estimates $\hat{\pmb{x}}$ will be required. We call this the 'Perfect SFDIA', that requires the attacker to be omnipotent and hence is impractical. The attacker thus needs to come up with innovative AC-SFDIA design methods with low information requirement.

\subsection{Adversarial attacks}\label{subsec:adv_attacks}
Adversarial attacks have been historically studied in the context of classification tasks like image classification, segmentation, speech recognition, etc \cite{Wiyatno2019AdversarialEI, NIPS2017_d494020f, 9063523, 9609659, Papernot2017PracticalBA}. In this proposed work, however, the adversary imitates the unknown AC-PSSE module using a substitute regression model trained using incomplete historical data from the power system and aims to generate adversarial examples with the substitute model as the target. Hence existing methods for adversarial attack design cannot be directly used in this case. The task of finding a suitable adversarial attack vector against a regression network, $g(\pmb{x}):\mathbb{R}^p \to \mathbb{R}^q$ is formulated as a constrained optimization problem as shown below:

\begin{align}
    & min_{\hspace{0.5mm}\pmb{\eta}} \hspace{5mm} ||\pmb{\eta}||_2 \label{eq:adv_opt_1_obj}\\
    & s.t.\hspace{5mm} ||g(\pmb{x}+\pmb{\eta}) – g(\pmb{x})||_2 \geq \rho \label{eq:adv_opt_1_con}
\end{align}
or equivalently,
\begin{align}
    & max_{\hspace{0.5mm}\pmb{\eta}} \hspace{5mm} ||g(\pmb{x}+\pmb{\eta}) – g(\pmb{x})||_2 \label{eq:adv_opt_2_obj}\\
    & s.t.\hspace{5mm} ||\pmb{\eta}||_2 \leq \epsilon \label{eq:adv_opt_2_con}
\end{align}

The optimization problem (\ref{eq:adv_opt_1_obj}, \ref{eq:adv_opt_1_con}) aims to find a perturbation, $\pmb{\eta}$, with minimal L2 norm such that the norm of the difference between the output of the regression network, $g$, before and after the perturbation is added to its input, $\pmb{x}$, is greater than a threshold, $\rho$. Equivalently, (\ref{eq:adv_opt_2_obj}, \ref{eq:adv_opt_2_con}) represents the dual formulation. In the subsequent sections, the dual formulation will be used. $\pmb{\eta}$ can then be called an adversarial perturbation against $g$. An immediate problem however is that the objective in Eq. \ref{eq:adv_opt_2_obj} is a highly non-convex function of $\pmb{\eta}$ due to the neural network $g$. In Section \ref{subsec:adv_opt}, a novel convex reformulation of (\ref{eq:adv_opt_2_obj}, \ref{eq:adv_opt_2_con}) is presented that requires significantly less computational effort to develop arbitrarily "good" adversarial perturbations.\par
Existing works, that propose adversarial attacks against learning based anomalous data detection models in the power system, design white box attacks explicitly targeting a particular defensive algorithm modelled as a classification network and hence are not effective against other defensive systems. 
DeeBBAA, the adversarial attack strategy proposed in this work, instead designs transfer based black box adversarial attacks against the AC-PSSE itself and the adversarial examples hence generated are highly transferable and can bypass a wide range of defenses, including conventional, statistical and learning based approaches.

\section{Proposed Attack Strategy}\label{sec:Proposed method}
This section consists of a detailed description of the proposed adversarial cum stealthy false data injection strategy using the DeeBBAA framework from the attacker's perspective. There are three main steps involved in the process:
\begin{enumerate}
    \item Identification of Attack Region and Reconnaissance
    \item Partial Approximation of the Unknown AC-PSSE
    \item Adversarial Optimization Against Proxy State Estimator
\end{enumerate}
These three steps implicitly conform to the two stage process involved in the design of transfer based black box attacks. The first two steps correspond to the collection of historical power system data and development of a substitute model mimicking the target AC-PSSE module which is carried out offline. Whereas the third step corresponds to the online design of an adversarial perturbation against real-time measurements using white box strategies on the substitute model. The following parts describe the three steps in further detail.
Before moving into further details, some assumptions about the resourcefulness of the attacker need to be made:\\
\textit{Assumption 1:} Once the attacker identifies a suitable attack region, it can collect historical power system data, including only power measurements and estimated states corresponding to the buses and lines included in the attack region. This constitutes the reconnaissance phase of the attacker.

This assumption on the capabilities of an attacker is a standard premise amongst the literature on data-driven SFDIA design \cite{9506908, 9853635, 9518372}. Note that unlike these works, historical data is required corresponding to only the sensors present in the attack region and not the entire power system for training the substitute regression network and no querying interaction is required between the attacker and the target state estimation module at any stage of the process. Historical data collection is required only once for training.

\subsection{Identification of Attack Region and Reconnaissance}\label{subsec:att_reg_id}
Prior to carrying out attacks and reconnaissance, the attacker identifies a suitable attack region that consists of a subset of the measurement units present in the power network. The attack region is characterized by a subset of buses, $\mathcal{B}_A \subset\ \mathcal{N}$ and a subset of branches $\mathcal{E}_A$ incident on the buses in the set $\mathcal{B}_A$ such that $\mathcal{E}_A\ \subset\ \mathcal{L}$. Each bus $b\ \in\ \mathcal{B}_A$, is equipped with SCADA measurement units providing the nodal real and reactive power injections $\{p_b, q_b\}$ and the voltage magnitude $|v_b|$. For designing the attack, however, the adversary needs to collect only the nodal power injections at a bus and the voltage magnitude is not required. Each line $l \in \mathcal{E}_A $ consists of SCADA measurement units measuring the sending and receiving end active and reactive power flows $\{p_l^s, q_l^s, p_l^r, q_l^r\}$. Thus, the attack region of the adversary is defined by the tuple of sets $\{\mathcal{B}_A,\mathcal{E}_A\}$. Depending on whether the subset of buses and lines chosen in the attack region form a connected induced subgraph of the power network or are random in nature, two types of attack regions are defined as described below. For illustration purposes, the graph structure of a standard IEEE 14 bus network is shown in Figures 3 and 4 where the nodes represent buses and the edges represent connecting power lines. 

\begin{figure}[!ht]
    \centering
    \includegraphics[ keepaspectratio,width=\columnwidth]{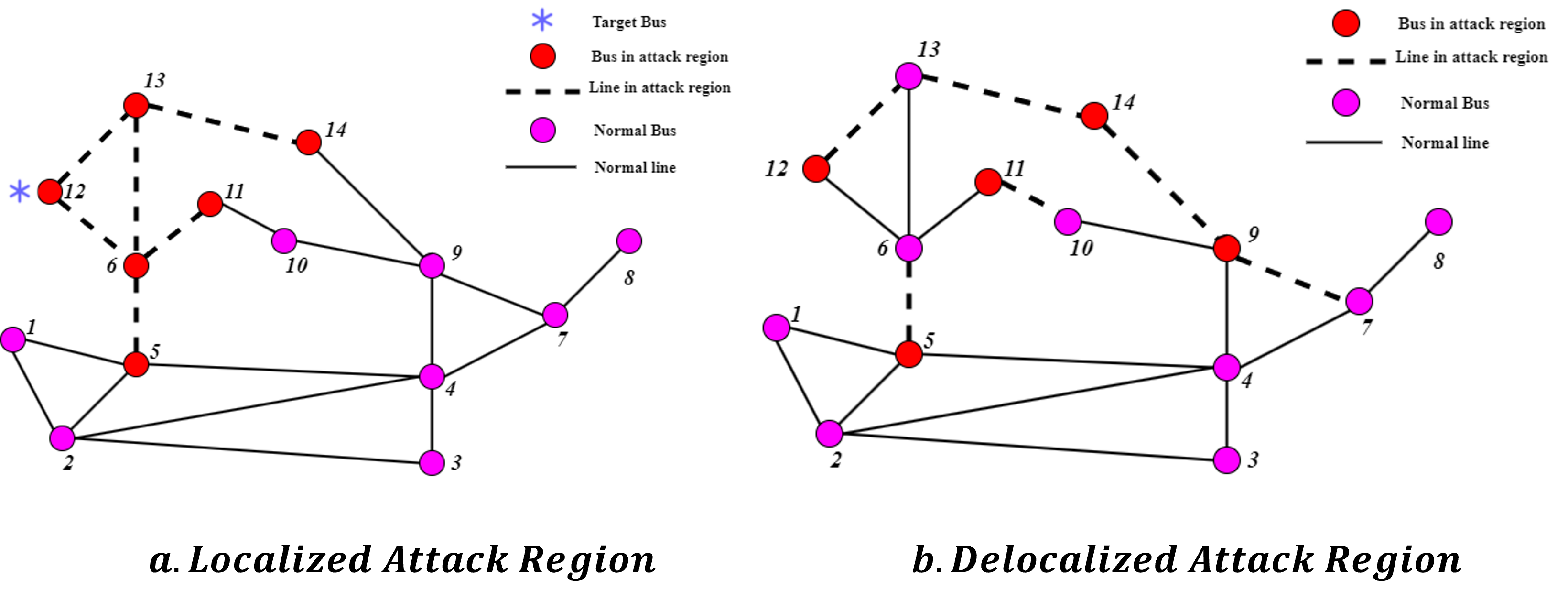}
    \caption{Graph representations of the IEEE 14 bus system demonstrating different types of attack regions. The red nodes represent buses in the attack region, the dotted edges indicate power lines included in the attack region, the magenta nodes and the solid edges represent buses and power lines not included in the attack region. The bus with the blue asterisk beside it represents the initial target bus for the localized case.}
    \label{fig:att_regs}
\end{figure}
\begin{enumerate}
    \item \textit{Localized Attack Region}: In this case, a target load bus is initially selected and all its k-hop neighboring non-generator buses including itself are included in $\mathcal{B}_A$. The lines connecting the buses in $\mathcal{B}_A$ are included in $\mathcal{E}_A$. Such attack regions have been widely considered in literature \cite{7401110, 8260948, 8425789, 7366616}. Figure \ref{fig:att_regs}.a depicts the formulation of a localized attack region on an IEEE 14 bus system. Bus 12 is selected as the initial target bus following which the 2-hop neighbors, i.e. the immediate neighbors and the immediate neighbors of immediate neighbors of bus 12 are included into the set of targeted buses, $\mathcal{B}_A$, i.e. $\mathcal{B}_A = \{5,6,11,12,13,14\}$. Finally all the lines connecting the buses in $\mathcal{B}_A$ are included into the set of target lines, $\mathcal{E}_A$. The attack region, $\{\mathcal{B}_A,\mathcal{E}_A\}$, thus formed represents a localized attack region.
    \item \textit{Delocalized Attack Region}: $|\mathcal{B}_A|$  number of non-generator buses are randomly selected from amongst the set of buses $\mathcal{N}$. Following this, a random subset of all the lines incident on the buses in $\mathcal{B}_A$ and whose other end is incident on a non-generator bus are included in $\mathcal{E}_A$. Figure \ref{fig:att_regs}.b depicts the formulation of a delocalized attack region on the IEEE 14 bus system. Buses $5,9,11,12$ and $14$ are randomly selected as the target buses, i.e., $\mathcal{B}_A = \{5,9,11,12,14\}$. For each of the buses in $\mathcal{B}_A$, a random subset of power lines incident on that bus and another non-generator bus are selected and included in $\mathcal{E}_A$. As an example, out of the four lines $\{(9,4), (9,7), (9,10), (9,14)\}$ incident on the targeted bus $9$, a random subset of $\{(9,7),(9,14)\}$ is selected to be included in $\mathcal{E}_A$. The same goes for all the other buses in $\mathcal{B}_A$. In the example shown in Fig. \ref{fig:att_regs}.b, $\mathcal{E}_A = \{(5,6),(9,7),(9,14),(11,10),(12,13)\}$.\\
    A delocalized attack region represents a much broader family of attack regions than a localized attack region which is a subset of the former. The intuition behind formulating a delocalized attack region is to include the following cases: \begin{itemize}
        \item The adversary doesn't have access to all measurement units from an electrically localized region of the power network.
        \item Some buses serving crucial loads may be more tightly protected than others by the Power System Operator. The adversary then chooses not to attack those buses.
    \end{itemize}
    These are some of the many possible use cases of formulating a delocalized attack region that provides a lot of flexibility for the adversary. 
\end{enumerate}

Once a suitable attack region is identified, the attacker performs reconnaissance. During that procedure it collects historical data corresponding to the attack region consisting of power injections at each bus, power flows in each line and state estimates at each bus present in the attack region. This data collection can be carried out either by eavesdropping in the communication network connecting the Remote Terminal Units to the SCADA system over a long period of time during which the attacker simply collects data and doesn't carry out any attacks \cite{9506908, 9853635, 9518372}.

\subsection{Partial Approximation of the Unknown AC-PSSE}\label{subsec:learn_nse}
\begin{figure}[!ht]
    \centering
    \includegraphics[ keepaspectratio,width=0.7\columnwidth]{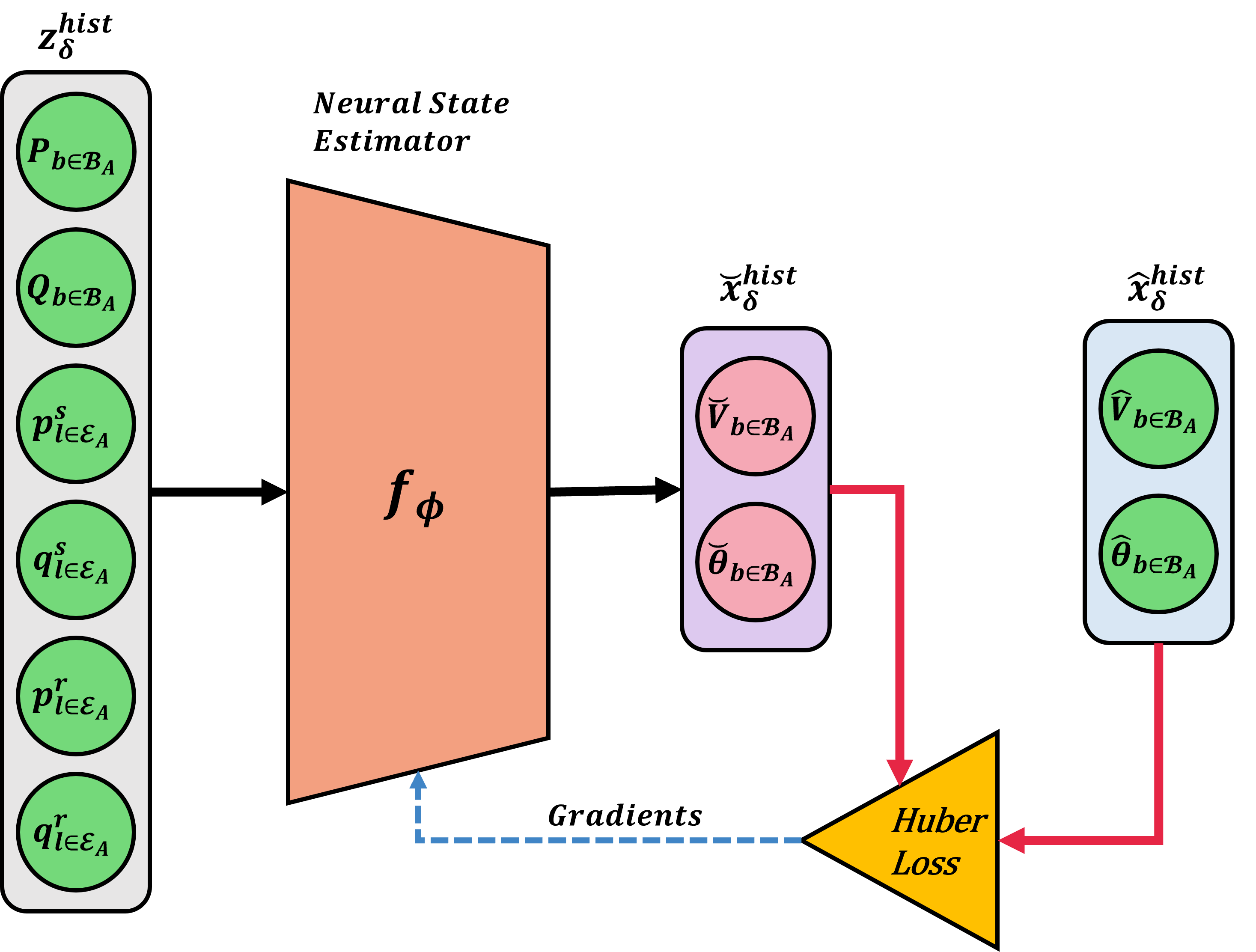}
    \caption{Training the Neural State Estimator $f_\phi$ using historical data samples.}
    \label{fig:nse_deebba}
\end{figure}

Having collected historical measurements and estimated states corresponding to the targeted attack region, the adversary trains a neural network, called the Neural State Estimator(NSE) to mimic the unknown AC-PSSE module. The input to the NSE are the historical power measurements corresponding to the identified attack region, i.e., the power injections at each bus and power flows at each line in the attack region. The target output are the historical estimated states corresponding to the attack region. Thus, with an attack region specified by the tuple ($\mathcal{B_A}, \mathcal{E_A}$), the input to the NSE has dimensions $(2\times \#(\mathcal{B}_A)+4\times \#(\mathcal{E}_A))$ (corresponding to 2 measurements per bus, i.e., active and reactive power injections and 4 measurements per line, i.e., sending and receiving end active and reactive power flows) and the output has dimensions $(2\times \#(\mathcal{B}_A))$(corresponding to 2 states per bus, i.e., estimated voltage magnitude and angle). Formally, the adversary learns a mapping $f_\phi : \mathcal{Z}_{\delta} \to \mathcal{X}_{\delta}$ such that
\begin{align}
    & \Breve{\pmb{x}}_{\delta}^{hist} = f_\phi(\pmb{z}_{\delta}^{hist})
\end{align}
where\\
$\pmb{z}_{\delta}^{hist} = [\{p_b, q_b\}_{b\in \mathcal{B}_A}, \{p_l^s, q_l^s, p_l^r, q_l^r\}_{l \in \mathcal{E}_A}] \in \mathbb{R}^{2\times \#(\mathcal{B}_A)+4\times \#(\mathcal{E}_A)}$ and $\hat{\pmb{x}}_{\delta}^{hist} = [\{\hat{|v_b|}, \hat{\theta}_b\}_{b \in \mathcal{B}_A}] \in \mathbb{R}^{2\times \#(\mathcal{B}_A)}$ are the partial historical power measurement and estimated state vectors corresponding to the attack region respectively. $\Breve{\pmb{x}}_{\delta}^{hist}$ is an estimate of $\hat{\pmb{x}}_{\delta}^{hist}$ obtained at the output of the NSE while $\mathcal{X}_{\delta}$ and $\mathcal{Z}_{\delta}$ are the spaces spanned by the partial historical estimated state and measurement vectors respectively. $f_\phi$ is the Neural State Estimator parameterized by $\phi$.
If $\mathcal{D} = \{\{\pmb{z}_{\delta}^{hist,i}, \hat{\pmb{x}}_{\delta}^{hist,i}\}_{\forall i = 1 to N_s}\}$ be the dataset of $N_s$ samples collected by the adversary during reconnaissance, then the trained neural state estimator $f_{\phi^*}$, parameterized by $\phi^*$, is obtained as follows:

\begin{align}
    & \phi^* = argmin_{\phi} \sum_{i=1}^{N_s} \mathcal{L} (\hat{\pmb{x}}_{\delta}^{hist,i} , f_\phi(\pmb{z}_{\delta}^{hist,i}))\\
    & \hspace{3.5mm} = argmin_{\phi} \sum_{i=1}^{N_s} \mathcal{L} (\hat{\pmb{x}}_{\delta}^{hist,i} , \Breve{\pmb{x}}_{\delta}^{hist,i})
\end{align}

where $\mathcal{L}(\pmb{a},\pmb{b})$ is an appropriate distance measure that quantifies the mismatch between two vectors $\pmb{a}$ and $\pmb{b}$.
Intuitively, the trained Neural State Estimator $f_{\phi^*}$ acts as a substitute partial state estimator learnt from the limited historical data available with the adversary from a small subset of measurement units in the power network. Figure \ref{fig:nse_deebba} provides a pictorial representation of the training procedure of the Neural State Estimator.
\\
The Neural State estimator used in this work comprises of a vanilla Multi Layer Perceptron (MLP) with two hidden layers, each consisting of 512 neurons. A Leaky ReLU activation function is used after each hidden layer to induce non-linearity. A dropout regularization is used after every hidden layer where each hidden neuron is dropped with a probability of 20\% during training. A Huber loss function with threshold $\gamma$ is used as the distance function, $\mathcal{L}$, to calculate the mismatch between the NSE output and the historical estimated state values corresponding to each training data sample. The Huber Loss for the $ i $-th data sample, $\{\pmb{z}_{\delta}^{hist, i}, \hat{\pmb{x}}_{\delta}^{hist, i}\}$, is given by:

\begin{align}
    \mathcal{L}_{\gamma}(\hat{\pmb{x}}_{\delta}^{hist, i} , \Breve{\pmb{x}}_{\delta}^{hist, i})=
    \begin{matrix}
        \frac{1}{2}||\Delta \pmb{x}_{hist, i}||_2^{2} \hspace{4mm} ,if \hspace{1mm} ||\Delta \pmb{x}_{hist, i}||_1  < \gamma\\
        \gamma (||\Delta \pmb{x}_{hist, i}||_1 - \frac1 2 \gamma) \hspace{4mm} ,otherwise
    \end{matrix}
\end{align}
where
$\Delta \pmb{x}_{hist, i} = \hat{\pmb{x}}_{\delta}^{hist, i} - \Breve{\pmb{x}}_{\delta}^{hist, i} =\hat{\pmb{x}}_{\delta}^{hist, i} - f_\phi(\pmb{z}_{\delta}^{hist, i})$ \\

The NSE is trained using backpropagation and the Adam optimizer in a minibatch fashion. For our work, we assume $\gamma = 1$. All the power measurements and voltage phasors are normalized respectively before training. 
In the next stage of the process, the adversary develops a black box adversarial attack on the learnt Neural State Estimator to generate joint stealthy false data and adversarial examples.

\subsection{Adversarial Optimization Against Proxy State Estimator}\label{subsec:adv_opt}

Having learnt the substitute model, $f_{\phi^*}$, for the unknown AC-PSSE using incomplete historical power system data, the adversary can now inject adversarial perturbations to real-time measurements in the attack region. For injecting stealthy adversarial examples, the attacker collects the current real-time power measurements from the attack region, solves a convex adversarial optimization problem to compute a suitable perturbation vector and adds it to the measurements before injecting them into the power network. Note that, at this online attack generation stage, the attacker only needs access to power measurements from the attack region, i.e., power injections and line flows. In the following part, a convex reformulation of the complex non-convex generic adversarial attack design framework for regression networks, given by equations (\ref{eq:adv_opt_2_obj}, \ref{eq:adv_opt_2_con}) as presented in Section \ref{subsec:adv_attacks} is carried out for the NSE, $f_{\phi^*}$.\par
Rewriting equations (\ref{eq:adv_opt_2_obj}, \ref{eq:adv_opt_2_con}) to include the Neural State Estimator, $f_{\phi^*}$, and its corresponding inputs as following:

\begin{align}
    & max_{\hspace{0.5mm}\pmb{\eta}} \hspace{5mm} ||f_{\phi^*}(\pmb{z}_{\delta}+\pmb{\eta}) – f_{\phi^*}(\pmb{z}_{\delta})||_2 \label{eq:adv_opt_3_obj}\\
    & s.t.\hspace{5mm} ||\pmb{\eta}||_2 \leq \epsilon \label{eq:adv_opt_3_con}
\end{align}

$\pmb{z}_{\delta}$ is the partial real-time power measurement vector collected from the attack region and $\pmb{\eta}$ is the desired adversarial perturbation. For convenience, the formulation in equations (\ref{eq:adv_opt_3_obj}, \ref{eq:adv_opt_3_con}) is renamed as the perturbation constrained deviation maximization (PCDM) problem. A two step relaxation strategy is hereby proposed to reduce the complexity of the PCDM problem and reformulate it as a convex optimization problem. First, a Taylor's Series approximation is carried out to the objective of the PCDM problem to convert it to a quadratic function following which a Semi-Definite Programming or SDP reformulation and subsequent convex relaxations are carried out in the second stage. The two stage relaxation strategy is explained below in greater detail:

\begin{enumerate}
\item \textit{Taylor’s first order approximation of the PCDM objective}:
\begin{align}\label{eq:tay_app}
    & f_{\phi^*}(\pmb{z}_{\delta}+\pmb{\eta}) = f_{\phi^*}(\pmb{z}_{\delta}) + \pmb{J}\pmb{\eta} 
\end{align}
where $\pmb{J}$ is the Jacobian matrix of $f_{\phi^*}$ with respect to its input $\pmb{z}_{\delta}$ calculated at $\pmb{z}_{\delta}$, i.e., $J_{ij} = \frac{\partial [f_{\phi^*}(\pmb{z}_{\delta})]_i}{\partial z_j}$ where $[f_{\phi^*}(\pmb{z}_{\delta})]_i$ is the $i$-th element of the vector $f_{\phi^*}(\pmb{z}_{\delta})$ and $z_j$ is the $j$-th element of $\pmb{z}_{\delta}$.
Replacing $f_{\phi^*}(\pmb{z}_{\delta}+\pmb{\eta})$ in Equation \ref{eq:adv_opt_3_obj} with its first order taylor’s approximation (\ref{eq:tay_app}), the PCDM problem (\ref{eq:adv_opt_3_obj}, \ref{eq:adv_opt_3_con}) becomes:

\begin{align}\label{eq:simp_QCQP_1}
    & max_{\hspace{0.5mm}\pmb{\eta}} \hspace{5mm} ||\pmb{J}\pmb{\eta}||_2\\
    & s.t. \hspace{5mm} ||\pmb{\eta}||_2 \leq \epsilon
\end{align}

or equivalently

\begin{align}
    & max_{\hspace{0.5mm}\pmb{\eta}} \hspace{5mm} \pmb{\eta}^T \pmb{J}^T \pmb{J} \pmb{\eta} \label{eq:simp_QCQP_2_obj}\\
    & s.t. \hspace{5mm} \pmb{\eta}^T \pmb{\eta} \leq \epsilon^2 \label{eq:simp_QCQP_2_con}
\end{align}

The superscript $T$ represents the transpose operation. The objective function is now simplified from a complex non-linear function consisting of a neural network to a much simpler quadratic function of $\pmb{\eta}$. The PCDM is thus relaxed into a Quadratic Constrained Quadratic Programming (QCQP) problem given by equations (\ref{eq:simp_QCQP_2_obj}, \ref{eq:simp_QCQP_2_con}). The solution of the QCQP will be a local optima of the original PCDM problem. However, QCQPs are also in general NP-hard and non-convex in nature. Thus, further relaxations are necessary to convert it into a convex problem.\\

\item \textit{Semi-Definite Programming Relaxation of QCQP}:

The QCQP (\ref{eq:simp_QCQP_2_obj}, \ref{eq:simp_QCQP_2_con}) can be reformulated into an SDP by making the following considerations:
\begin{align}
    & \pmb{J}^T\pmb{J} = \pmb{\tilde{J}} \label{eq:jTj}\\
    & \pmb{\eta} \pmb{\eta}^T = \pmb{W} \label{eq:nnT_eq_W}
\end{align}
Note that,
\begin{equation}\label{eq:res_trace}
    \pmb{\eta}^T \pmb{\tilde{J}} \pmb{\eta} = Tr(\pmb{\tilde{J}}\pmb{\eta}\pmb{\eta}^T) = Tr(\pmb{\tilde{J}} \pmb{W})
\end{equation}

Here $Tr(\pmb{A})$ represents the trace or the sum of diagonal elements of matrix $\pmb{A}$. Using result (\ref{eq:res_trace}), the QCQP, (\ref{eq:simp_QCQP_2_obj}, \ref{eq:simp_QCQP_2_con}), can be converted to the SDP shown below:  

\begin{align}
    & max_{\hspace{0.5mm}\pmb{W}} \hspace{5mm} Tr(\pmb{\tilde{J}} \pmb{W}) \label{eq:SDP_exact_obj}\\
    & s.t.\hspace{5mm} \pmb{W} \succeq 0 \label{eq:SDP_exact_con1}\\
    & \hspace{10mm} Tr(\pmb{W}) \leq \epsilon^2 \label{eq:SDP_exact_con2}\\
    & \hspace{10mm} Rank(\pmb{W}) = 1 \label{eq:SDP_exact_con3}
\end{align}

Eq. \ref{eq:SDP_exact_obj} is derived from result (\ref{eq:res_trace}). Constraint (\ref{eq:SDP_exact_con1}) enforces the condition that $\pmb{W}$ is a positive semi-definite matrix. Constraint (\ref{eq:SDP_exact_con2}) is obtained by applying $\pmb{\eta}^T \pmb{\eta} = Tr(\pmb{\eta} \pmb{\eta}^T) = Tr(\pmb{W})$ and Eq. \ref{eq:nnT_eq_W} to constraint \ref{eq:simp_QCQP_2_con}. Constraint \ref{eq:SDP_exact_con3} is a direct consequence of Eq. \ref{eq:nnT_eq_W} where $\pmb{\eta}$ is a vector. Note that the presence of the rank constraint \ref{eq:SDP_exact_con3} renders this SDP non-convex and NP-hard. However, if a rank one matrix $\pmb{W}$ can be obtained as the solution to the SDP (\ref{eq:SDP_exact_obj})-(\ref{eq:SDP_exact_con3}) then $\pmb{W}$ uniquely and exactly solves the QCQP given by equations (\ref{eq:simp_QCQP_2_obj})-(\ref{eq:simp_QCQP_2_con}). For convenience, the SDP formulation given by equations (\ref{eq:SDP_exact_obj} - \ref{eq:SDP_exact_con3}) is named as the exact SDP formulation of the QCQP, (\ref{eq:simp_QCQP_2_obj},\ref{eq:simp_QCQP_2_con}). \\
A final convex relaxation is carried out to convert the non-convex rank equality to a convex inequality by replacing the rank function by its closest convex approximation, the nuclear norm. The nuclear norm of a matrix $\pmb{W}$ is defined as the sum of its singular values, i.e., $||\pmb{W}||_* = Tr(\sqrt{\pmb{W}^T\pmb{W}})$. The final convex SDP reformulation of the PCDM problem, (\ref{eq:adv_opt_2_obj}, \ref{eq:adv_opt_2_con}),  is thus given by:
\begin{align}
    & max_{\hspace{0.5mm}\pmb{W}} \hspace{5mm} Tr(\pmb{\tilde{J}} \pmb{W}) \label{eq:SDP_conv_obj}\\
    & s.t.\hspace{5mm} \pmb{W} \succeq 0 \label{eq:SDP_conv_con1}\\
    & \hspace{10mm} Tr(\pmb{W}) \leq \epsilon^2 \label{eq:SDP_conv_con2}\\
    & \hspace{10mm} ||\pmb{W}||_* \leq 1 \label{eq:SDP_conv_con3}
\end{align}

\textit{Result 1}: If the principle eigenvalue of the matrix $\pmb{W}$ obtained after solving the convex SDP represented by equations (\ref{eq:SDP_exact_obj}-\ref{eq:SDP_exact_con3}) is infinitesimally close to one, then it can be concluded that $\pmb{W}$ is an approximately rank one matrix. \par

\textit{Proof:} Since $\pmb{W}$ is a positive semi-definite matrix, all its eigenvalues are non-negative. Suppose $\lambda^* = \lambda_1 \geq \lambda_2 \geq ... \geq \lambda_n \geq 0$ be the $n$ eigenvalues (in descending order of absolute values) of $\pmb{W} \in \mathbb{R}^{n*n}$ and $\lambda^*$ be the principle eigenvalue. Let $\lambda^* = 1-\delta$, where $\delta$ is an arbitrarily small non-negative number close to zero.\\
According to Equation \ref{eq:SDP_conv_con3}, 
\begin{align}
    & ||\pmb{W}||_* \leq 1\\
    & \implies Tr(\sqrt{\pmb{W}^T\pmb{W}}) \leq 1\\
    & \implies \sum_{i=1}^{n} \lambda_i \leq 1 \\
    & \implies (1-\delta) + \sum_{i=2}^{n} \lambda_i \leq 1\\
    & \implies \sum_{i=2}^{n} \lambda_i \leq \delta \\
    & \implies \lim_{\delta\to 0} \sum_{i=2}^{n} \lambda_i \to 0\\
    & \because \lambda_i \geq 0 \hspace{2mm}\forall i \implies \lambda_i \to 0 \hspace{2mm}\forall i=2\hspace{1mm} to\hspace{1mm} n
\end{align}

Result 1 states that if the principle eigenvalue of $\pmb{W}$ infinitesimally close to 1 then the remaining eigenvalues are very close to zero, implying that $\pmb{W}$ is an approximately rank one matrix. Under these conditions the $\pmb{\eta}$ derived from $\pmb{W}$ would be a solution arbitrarily close to a local optima of the original PCDM problem. Empirical evidence is provided in the supplementary document establishing that the dominant eigenvalue of $\pmb{W}$ obtained after solving the convex SDP formulation is infinitesimally close to 1 and the second largest eigenvalue is infinitesimally close to 0.
\end{enumerate}
\begin{figure}[!ht]
    \centering
    \includegraphics[ keepaspectratio,width=\columnwidth]{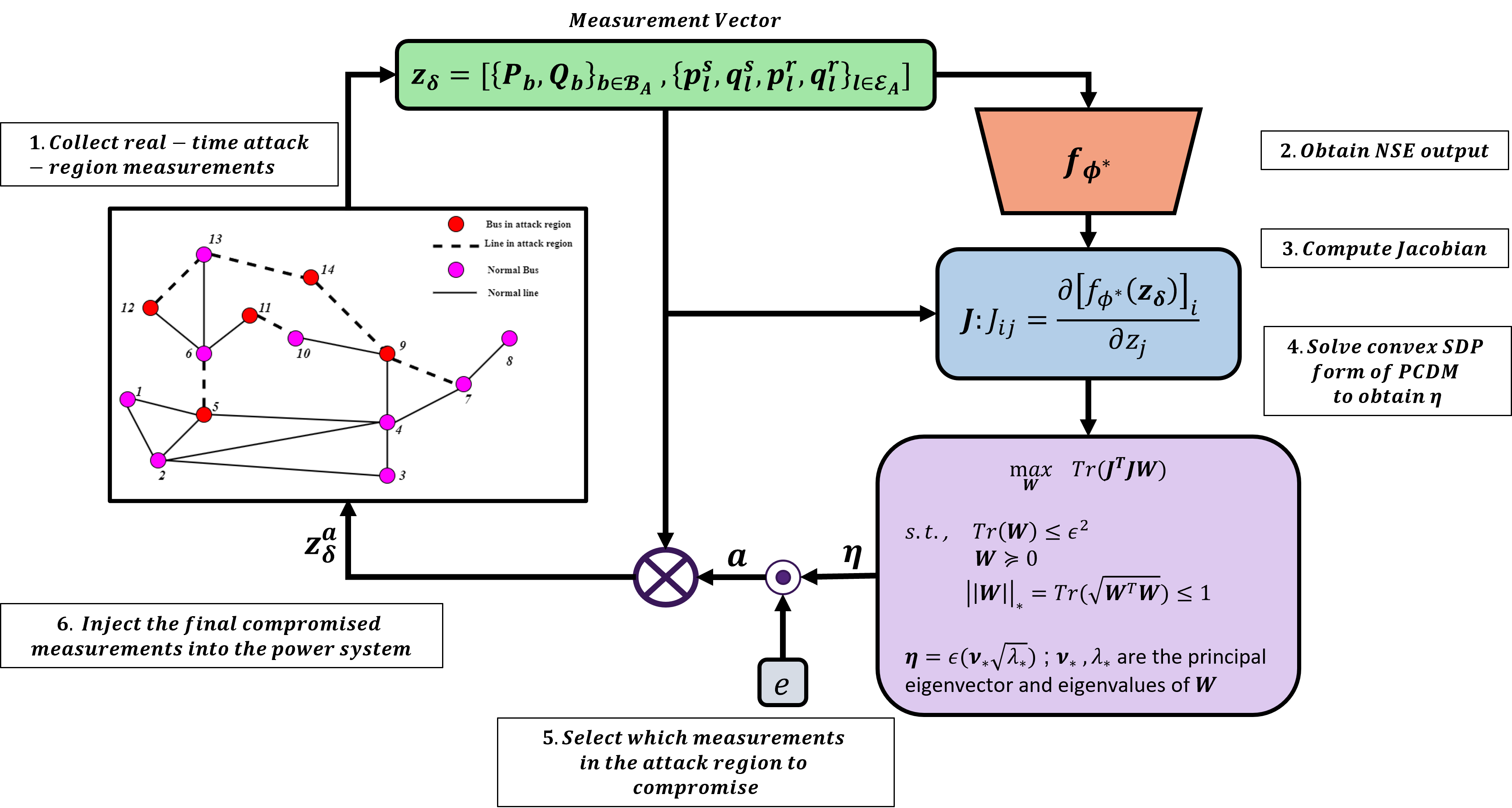}
    \caption{Stealthy Black Box Adversarial Attack Formulation against Power System State Estimation using the proposed DeeBBAA framework.}
    \label{fig:deebbaa}
\end{figure}
After obtaining $\pmb{W}$ by solving the convex SDP, the adversary computes its principle eigenvalue, $\lambda^*$, and principal eigenvector, $\pmb{\nu}^*$, using singular value decomposition. The principal eigenvector is the orthonormal eigenvector corresponding to the principle eigenvalue of $\pmb{W}$. The optimal adversarial perturbation, $\pmb{\eta}$, and the compromised measurement vector $\pmb{z}_{\delta}^a$ are then derived as follows:
\begin{align}
    & \pmb{\eta} = \epsilon \sqrt{\lambda^*} \pmb{\nu}^* \label{eq:compute_eta}\\
    & \pmb{z}_{\delta}^a = \pmb{z}_{\delta} + \pmb{\eta} \odot \pmb{e} \label{eq:compute_za}
\end{align}
where $\odot$ represents the element-wise product or the Hadamard product and $\pmb{e}$ is the compromised measurement selection vector, having the same dimensions as $\pmb{z}_{\delta}$. The $i$-th element of $\pmb{e}$ can be either $1$ or $0$ depending on whether the adversary wishes to inject adversarial perturbation to the $i$-th measurement contained in $\pmb{z}_{\delta}$ or not. This provides an additional level of flexibility to the adversary in terms of attack sparsity.\par
Figure \ref{fig:deebbaa} shows a block diagram depicting the formulation of a stealthy black box adversarial attack against power system state estimation using the DeeBBAA framework. For attacking the power system at a particular time instant $t$, the adversary collects power measurements from the attack region, following which the trained Neural State Estimator,$f_{\phi^*}$ is used to compute the Jacobian matrix $\pmb{J}$ using backpropagation. The convex SDP problem defined by Equations (\ref{eq:SDP_exact_obj}-\ref{eq:SDP_exact_con3}) is then solved to obtain an approximately rank-one matrix $\pmb{W}$ from which the optimal adversarial perturbation and input is computed according to Equations \ref{eq:compute_eta} and \ref{eq:compute_za}.

\section{Simulation and Results}\label{sec:case}
\subsection{Test System and Data Generation}\label{subsec:exp_test_sys}
\begin{figure}[!ht]
    \centering
    \includegraphics[ keepaspectratio,width=\columnwidth]{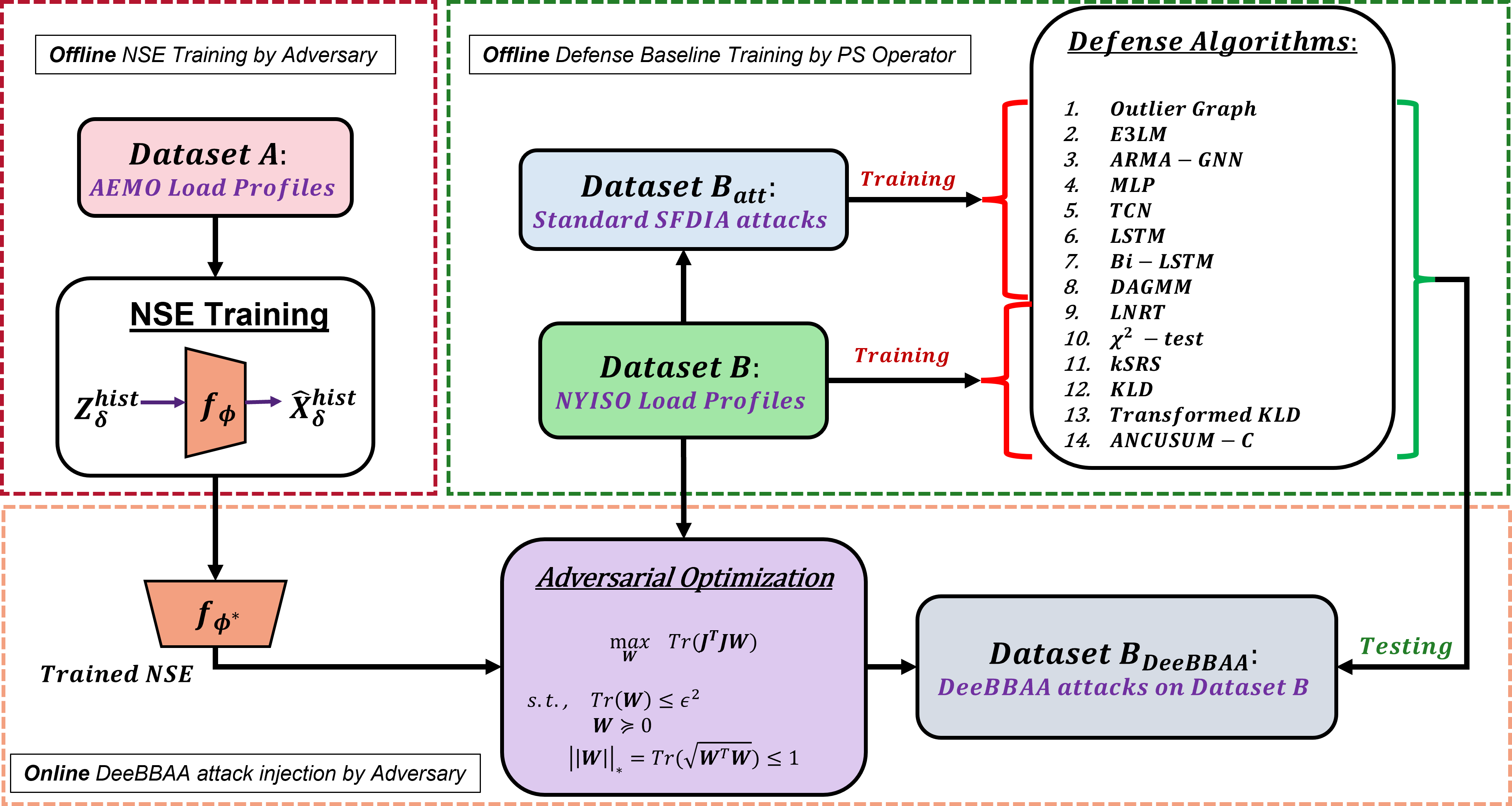}
    \caption{Summary of data generation and subsequent purpose of usage of different datasets. Dataset A is only used to train the NSE in the offline stage of DeeBBAA. Dataset B is used for training the defense baselines and online generation of attack vectors using DeeBBAA. This is done to ensure the worst case scenario from the attacker's perspective where the defense baselines are trained on the same data which is used to generate the online DeeBBAA attacks. At the same time, the NSE is trained on a completely different dataset to prevent data spillage.}
    \label{fig:deebbaa_data_usage}
\end{figure}

Due to unavailability of a dataset pertaining to the application of Stealthy False Data Injection in cyber-physical power systems, data is generated via simulations on the standard IEEE 39 bus and 118 bus test systems. These test systems are explained in detail in the Supplementary Material. The simulations were done in two stages: 
\begin{itemize}
    \item In the first stage, data representing normal operating conditions of the power system is generated. Two different real life load profiles corresponding to publicly available databases of the Australian Energy Market Operator (AEMO)\cite{AEMO} for the year of 2019 and the New York Independent System Operator (NYISO)\cite{NYISO}for the year of 2021 were selected. The AEMO load profiles have an interval length of 30 min while the NYISO load profile has a 5-minute interval span. The original AEMO load profiles are interpolated to a 5-minute interval span similar to the NYISO profiles. Following this the data generation algorithm proposed in \cite{9559412} is employed to obtain two datasets, each consisting of power system measurements and estimated states under normal operating conditions. The dataset generated from the AEMO load profiles, called Dataset A is used to train the Neural State Estimator, whereas the dataset generated from the NYISO load profiles, called Dataset B is used to train the defensive BDD algorithms and for testing the evasive properties of the DeeBBAA attacks. Both Dataset A and B have 105120 samples, each sample consisting of a tuple ($\pmb{z},\hat{\pmb{x}}$), where $\pmb{z} \in \mathbb{R}^{2\times \#(\mathcal{N})+4\times \#(\mathcal{M})}$ and $\hat{\pmb{x}} \in \mathbb{R}^{2\times \#(\mathcal{N})}$ represent the power measurements and corresponding estimated states. Note that voltage measurements are not included in $\pmb{z}$ as it is not required for attack formulation using DeeBBAA. However, voltage measurements are included for state estimation and training the defense algorithms.
    
    \item In the second stage, Dataset B is used to generate another dataset consisting of standard SFDIA attacks, which will henceforth be called Dataset $B_{att}$ for convenience. Half of the data samples in Dataset B are randomly selected for standard SFDI attack generation. Three conventional SFDIA generation strategies, used in \cite{9676996}, are used to generate the attack data samples. The SFDIA attack vector $\pmb{a}$ is generated using the generic SFDIA formulation given by equation \ref{eq:SFDIA}. Recall that the attacker needs both power system parameters like admittances and topology information (reflected in the realization of $h(.)$) and real-time state estimates, $\hat{\pmb{x}}$ to formulate the attack using equation \ref{eq:SFDIA} and the three strategies vary depending on whether the information residing with the attacker is perfect or noisy. The first conventional type of attack assumes that the attacker has perfect knowledge of the power system parameters and real-time state estimates and then uses equation \ref{eq:SFDIA} to formulate the attack vector. The second strategy assumes that the attacks are generated using imperfect power system parameter values, i.e., the power line parameters like admittance values are perturbed by adding 10\% random gaussian noise to them. In the third attack strategy, the attacker is assumed to have noisy estimates of the real-time state values with error levels varying between -8\% to 8\%. The vector $\pmb{c}$ is formulated randomly to consist of 10\% to 80\% non-zero elements whose values range between 10\% to 190\% of the original state values. A data sample from Dataset B is randomly selected with 50\% probability for attack generation and one of the three standard attack strategies is randomly selected with uniform probability for injecting false data into the selected sample. 
\end{itemize}
Figure \ref{fig:deebbaa_data_usage} summarizes the data generation and usage procedure for all the experiments carried out in this work.

\subsection{Training the Neural State Estimator}\label{subsec:exp_train_nse}
The Neural State Estimator described in Section \ref{subsec:learn_nse} is trained using Dataset A. Appropriate localized and delocalized attack regions, denoted by $(\mathcal{B_A},\mathcal{E_A})^{L}$ and $(\mathcal{B_A},\mathcal{E_A})^{D}$ are selected for both the IEEE test systems. The localized attack regions created for the IEEE 39 bus system consists of 8 buses and 8 power lines and that for the IEEE 118 bus system consists of 28 buses and 17 power lines. The delocalized attack regions for the IEEE 39 bus system consists of 7 buses and 10 power lines and that for the IEEE 118 bus system consists of 30 buses and 34 lines. The details of the buses and the power lines selected for each attack region can be found in the Supplementary material.

Dataset A is randomly split into a training and testing subsets with a 80-20 split to train the NSE. For each data sample in Dataset A, power measurements corresponding to the buses and lines in the attack region serve as input to the NSE and the estimated states corresponding to the buses in the attack region serve as the target output. Since for each of the two IEEE test systems, two attack regions have been defined, 4 different NSE models each corresponding to a test system and a particular attack region were needed to be trained. Training was carried out in a mini-batch fashion with batch a size of 256 for 50000 steps.

\subsection{Training the Defense Baselines}\label{subsec:exp_defense}

\begin{table*}[!ht]
\begin{center}
\caption{ Details of defensive baselines used to test the evasive properties of the SFDIA generated using DeeBBAA.
}
\label{tab:att_baselines}
\resizebox{\textwidth}{!}{
\renewcommand{\arraystretch}{1.2}
\begin{tabular}{|c|l|l|l|l|}
\hline
\textbf{Sl. No.} & \multicolumn{1}{c|}{\textbf{Name}}                                                                            & \multicolumn{1}{c|}{\textbf{Method Type}}                       & \multicolumn{1}{c|}{\textbf{Description}}                                                                                                                                                                                                                                                                                                                                                                                                                                                                                                & \multicolumn{1}{c|}{\textbf{Training Entails}}                                                              \\ \hline
1                & \begin{tabular}[c]{@{}l@{}}Largest Normalized Residue Test \\ (LNRT)\end{tabular}                             & \multirow{2}{*}{Conventional Residue based}                     & Hypothesis testing involves comparing the $L_{\infty}$ norm of the normalized measurement residue vector against a pre-computed threshold.                                                                                                                                                                                                                                                                                                                                                                               & Threshold computation                                                                                       \\ \cline{1-2} \cline{4-5} 
2                & Chi-Squared Test                                                                                              &                                                                 & Hypothesis testing involves comparing the $L_2$ norm of the normalized measurement residue vector against a pre-computed threshold.                                                                                                                                                                                                                                                                                                                                                                                                   & Threshold computation                                                                                       \\ \hline
3                & KL Divergence (KLD) Test \cite{7035067}                                                                                      & \multirow{5}{*}{Statistical Consistency Checks}                 & \begin{tabular}[c]{@{}l@{}}Compares the value of the KL Divergence between the empirical distributions of historical measurement residues and that of the current \\ measurement residues against a pre-cpmouted threshold.\end{tabular}                                                                                                                                                                                                                                                                                                 & Threshold computation                                                                                       \\ \cline{1-2} \cline{4-5} 
4                & Transformed KLD Test \cite{7961272}                                                                                         &                                                                 & Same as KLD test. KLD is computed after transforming the residues with a non-linear function like the power function.                                                                                                                                                                                                                                                                                                                                                                                                                    & Threshold computation                                                                                       \\ \cline{1-2} \cline{4-5} 
5                & \begin{tabular}[c]{@{}l@{}}k-Smallest Residue Similarity \\ (kSRS) Test \cite{9676996}\end{tabular}                          &                                                                 & \begin{tabular}[c]{@{}l@{}}Instead of checking the statistical consistency of measurement values, checks the consistency of the measurement residuals. Compute \\ one-time step deviation in measurement residual vectors over a reference interval and for each measurement channel compute the \\ Jensen Shannon Divergence between the empirical distribution of the deviation in current residuals and historical ones. The average of \\ the greatest k JSD values are then compared against a pre-computed threshold.\end{tabular} & Threshold computation                                                                                       \\ \cline{1-2} \cline{4-5} 
6                & \begin{tabular}[c]{@{}l@{}}Adaptive Non-linear Cumulative \\ Sum (ANCUSUM-C)\cite{9311640}\end{tabular}                     &                                                                 & \begin{tabular}[c]{@{}l@{}}Real time detection of FDIA by computing a cumulative sum of measurement residues using states estimated by exponential smoothing. \\ The paarmeters of the algorithm are adaptively modified and a decision statistic is calculated recursively for each measurement channel \\ and compared against a threshold to detect attack.\end{tabular}                                                                                                                                                              & Threshold computation                                                                                       \\ \cline{1-2} \cline{4-5} 
7                & Outlier Graph Based BDD \cite{9107408}                                                                                       &                                                                 & \begin{tabular}[c]{@{}l@{}}One time-step difference between measurement vectors are computed and an outlier graph is formed based on a 3-sigma test and the \\ topological placement of the measurements in the power network. The current measurement vector is considered anomalous if the number \\ of nodes in the outlier graph with degree greater than a particular threshold is greater than a precomputed value.\end{tabular}                                                                                                   & Threshold computation                                                                                       \\ \hline
8                & Multi-Layer Perceptron (MLP) \cite{9622117, 8334607}                                                                                 & Supervised point classification                                 & \begin{tabular}[c]{@{}l@{}}MLP is used to solve a supervised binary classification task with power system measurements as input. The output is a binary label \\ indicating whether the measurement integrity is compromised or not.\end{tabular}                                                                                                                                                                                                                                                                                        & Trained MLP weights                                                                                         \\ \hline
9                & Long Short Term Memory (LSTM) \cite{9773032, 8334607, 9542963}                                                                                & \multirow{3}{*}{Supervised sequence to sequence classification} & \multirow{3}{*}{\begin{tabular}[c]{@{}l@{}}Formulates SFDIA detection as a sequence to sequence classification task, where the input is a time series of measurement vectors \\ and the output is a sequence of binary labels indicating whether the measurements at a particular timestep is compromised or not.\end{tabular}}                                                                                                                                                                                                          & Trained LSTM weights                                                                                        \\ \cline{1-2} \cline{5-5} 
10               & Bidirectional LSTM (Bi-LSTM)                                                                                  &                                                                 &                                                                                                                                                                                                                                                                                                                                                                                                                                                                                                                                          & Trained Bi-LSTM weights                                                                                     \\ \cline{1-2} \cline{5-5} 
11               & \begin{tabular}[c]{@{}l@{}}Temporal Convolutional Network \\ (TCN) \cite{9557319, 9049087, 8233155}\end{tabular}                               &                                                                 &                                                                                                                                                                                                                                                                                                                                                                                                                                                                                                                                          & Trained TCN weights                                                                                         \\ \hline
12               & \begin{tabular}[c]{@{}l@{}}AutoRegressive Moving Average - \\ Graph Neural Network \\ (ARMA-GNN) \cite{9559412}\end{tabular} & \multirow{2}{*}{Supervised Localization}                        & \begin{tabular}[c]{@{}l@{}}Using the power network as the underlying graph an ARMA-GNN is trained on a localization task to identify which nodal states are \\ compromised.\end{tabular}                                                                                                                                                                                                                                                                                                                                                 & Trained ARMA-GNN weights                                                                                    \\ \cline{1-2} \cline{4-5} 
13               & \begin{tabular}[c]{@{}l@{}}Enhanced Ensemble of Extreme \\ Learning Machines (E3LM) \cite{9055170}\end{tabular}              &                                                                 & \begin{tabular}[c]{@{}l@{}}Ensemble of Extreme Learning Machines initialized using a gaussian random distribution and Latin Hypercube Sampling used for solving \\ a localization task where for each data sample, a binary classification task is solved for each of the nodes to indicate which nodes are \\ compromised.\end{tabular}                                                                                                                                                                                                 & \begin{tabular}[c]{@{}l@{}}Trained weights of the final layer \\ of ELMs\end{tabular}                       \\ \hline
14               & \begin{tabular}[c]{@{}l@{}}Deep Autoregressive Gaussian \\ Mixture Model (DAGMM) \cite{9706368}\end{tabular}                 & Unsupervised Energy based anomaly detection                     & \begin{tabular}[c]{@{}l@{}}A deep autoencoder is trained on the measurement vectors to obtain compresssed latent features from the encoder which is used along with \\ the reconstruction errors to fit a Gaussian Mixture Model. Using the parameters of the trained GMM and the encoder output, an energy function \\ is computed for each data sample whose outcome is compared against a pre-specified threshold to detect anomalous data.\end{tabular}                                                                              & \begin{tabular}[c]{@{}l@{}}Trained Autoencoder weights, \\ GMM parameters and energy threshold\end{tabular} \\ \hline
\end{tabular}
}\end{center}
\end{table*}

The false data generated using the DeeBBAA framework is tested against three broad types of state of the art defensive BDD algorithms existing in literature. The probability of the SFDIA bypassing these defensive baselines is computed as the number of attacked samples misclassified as benign to the total number of attacked samples in the test data. 14 BDD algorithms each belonging to one of three classes - conventional, statistical and learning based are considered. The conventional baselines include the residue-based BDD algorithms described in Section II.C, namely, the Largest Normalized Residue test and the Chi Squared test. The second group consists of detection strategies that check for consistency in the measurement data reflected by statistical measures like KL Divergence between historical and current residues\cite{7035067,7961272, 9676996}, normalized cumulative sum of residues \cite{9311640}, and structural properties of residual outlier graphs\cite{9107408} to detect the presence of anomalous data. The third group consists of data-driven learning based algorithms including supervised and unsupervised deep learning and extreme learning methods for classification of compromised data using power system measurements as inputs. The data-driven algorithms formulate the task of SFDIA detection either as a binary classification task in which the status of the overall power system is returned as either compromised or not compromised or as a localization task where for each of the possible nodes in the power system, a binary classification task is solved to identify the compromised locations. The details of the defensive algorithms are given in Table \ref{tab:att_baselines}. Other than the two conventional methods, 5 statistical and seven data-driven algorithms are considered, out of which 6 are supervised and one is unsupervised. Their training is carried out according to the procedure given in the original works that proposed them. Training the conventional and statistical methods entail the computation of thresholds of statistical measures to carry out hypothesis testing. Dataset B is used for this purpose, and a false alarm rate of 2\% is considered. Data-driven methods are trained to correctly distinguish benign measurement samples from those compromised using the standard SFDI attacks contained in Dataset $B_{att}$ as described in Section \ref{subsec:exp_test_sys}. Further details about the training procedure and model architectures have been shifted to the Supplementary document due to a lack of space.

\subsection{Designing attack vectors using DeeBBAA}\label{subsec:exp_deebbaa_att_design}

A randomly selected subset of one third data samples from Dataset B is used to inject false data using DeeBBAA. The dataset hence generated will be called Dataset $B_{DeeBBAA}$ for convenience. For each of the IEEE test systems considered in this work, adversarial attacks using four values of $\epsilon$- 1,2,5 and 10 - are carried out. For each of these cases, 3 different measurement selection vectors, $\pmb{e}$ are tested. The measurement selection vector $\pmb{e}$ has been introduced in Equation \ref{eq:compute_za}. The three values of $\pmb{e}$ vary in terms of the number of ones contained in $\pmb{e}$. The three cases of $\pmb{e}$ correspond to the following scenarios:
\begin{enumerate}
    \item all measurements from the attack region are injected with false data, i.e., $\pmb{e} = \pmb{1}^T$. If $N_A$ is the total number of measurements from the attack region, then $\pmb{e}^T\pmb{1} = N_A$ where $\pmb{1}^T$ is the vector of all ones. For the IEEE 39 bus system, $N_A$ is equal to 48 for the localized attack region case and equal to 54 for the delocalized attack region case. For the IEEE 118 bus system, $N_A$ is equal to 124 for the localized attack region case and equal to 200 for the delocalized attack region case.
    
    \item one half of the measurements from the attack region are injected with false data, i.e., the number of ones in $\pmb{e}$ is exactly equal to half its length. The adversary chooses to attack $N_A/2$ measurements which correspond to the greatest $N_A/2$ elements of $\pmb{\eta}$ in absolute terms. More formally, let $\pmb{r}$ be an integer vector consisting of $N_A$ unique indices from $1$ to $N_A$ such that $|\eta_{r_i}| \geq |\eta_{r_j}|$ if and only if $i<j$. $r_i$ is the $i$-th element of $\pmb{r}$ and $\eta_{r_i}$ is the element of $\pmb{\eta}$ corresponding to the index represented by the $i$-th element of $\pmb{r}$. Then, the $i$-th element of $\pmb{e}$, $e_i$, will be equal to 1 if and only if $i \in \mathcal{S}: \mathcal{S}=\{k:k=r_j \hspace{3mm}iff\hspace{3mm} j\leq N_A/2\}$, otherwise zero.
    
    \item a tenth of the measurements are injected with false data, i.e., the number of ones in $\pmb{e}$ is equal to one tenth of the total number of measurements obtained from the attack region. Again the adversary finds the greatest $N_A/10$ (rounded off to the nearest integer) elements in $\pmb{\eta}$ and makes the corresponding elements of $\pmb{e}$ equal to one while all the remaining elements of $\pmb{e}$ are made zero.
    
\end{enumerate}
For each of the IEEE test systems considered in this work, attacks are generated for each of the two choices of attack regions - localized and delocalized, and for each attack region 3 choices of the measurement selection vector, $\pmb{e}$, are explored as explained above, thus leading to 6 scenarios per IEEE test system. Corresponding to each scenario, different versions of Dataset $B_{DeeBBAA}$ is generated. For each of these scenarios, the probability of the data samples representing DeeBBAA attacks that successfully bypass the aforementioned defense baselines and the deviations in estimated states and power measurements induced by these attacks are analysed. The probability of successfully bypassing the defense baselines is calculated as the ratio of the number of compromised data samples that are wrongly classified as good data to the number of compromised data samples. The following section consists of the aforementioned results and the corresponding inferences.

\subsection{Results and Inferences}\label{subsec:exp_results}
In this section, the evasive properties of the DeeBBAA attacks is analysed for the defense baselines outlined in Section \ref{subsec:exp_defense} for different test systems, attack magnitudes and spread. Following this, the deviations introduced by the DeeBBAA attacks in the measurements and the corresponding state estimates is analysed.

\subsubsection{Bypassing Conventional BDD safeguards}\label{subsec:res_BDD}
\begin{figure}[!ht]
    \centering
    \includegraphics[keepaspectratio, width=1.02\columnwidth]{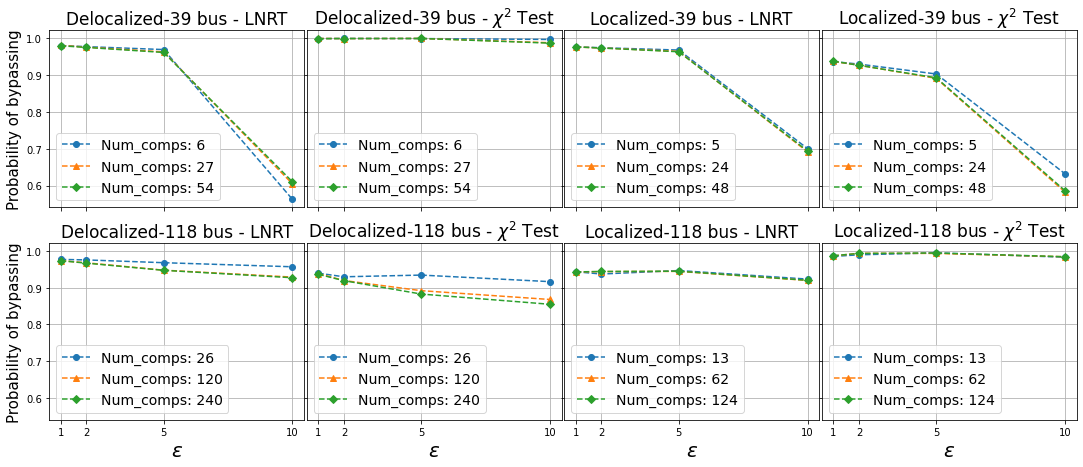}
    \caption{Probability of DeeBBAA attacks bypassing conventional LNRT and $\chi^2$ tests for IEEE 39 bus and 118 bus systems under different attack regions, epsilons and number of compromised measurements.}
    \label{fig:conv_defense_comp}
\end{figure}

In Figure \ref{fig:conv_defense_comp}, the probability of the DeeBBAA attacks bypassing the conventional BDD safegaurds, i.e., LNRT and the $\chi^2$ test, are plotted for the IEEE 39 bus and 118 bus test systems under different attack region types. The title of each subplot consists of three parts corresponding to the attack region type, the IEEE test system and the conventional BDD algorithm considered in that specific plot in the same order. In each subplot, the probability of bypassing BDD safeguards is compared against different values of epsilon and the total number of measurements compromised by the attacker, which is decided using the methodology outlined in Section \ref{subsec:exp_deebbaa_att_design}. Following are the observations and inferences related to this result:
\begin{enumerate}
    \item In general, DeeBBAA attacks generated on the IEEE 39 bus test system have a lower probability of bypassing the BDD when $\epsilon =10$ as compared to those generated on the IEEE 118 bus case.
    \item The LNRT is able to detect DeeBBAA attacks with maximum probabilities of 40\% and 30\% for the IEEE 39 bus case in the delocalized and localized attack region settings respectively only when  $\epsilon = 10$. Similarly, the $\chi^2$ test is able to detect localized attacks on the 39 bus system with a maximum probability of around 40\% when $\epsilon = 10$. In all the other cases, the probability of bypassing the BDD is at least 90\%.   
    \item As the value of $\epsilon$ increases, the magnitude of attack vectors increases proportionally and so does the deviation caused by them in the estimated state and measurement vectors. As a result attacks carried out with $\epsilon = 10$ is the most susceptible to detection. \item In most all the cases, LNRT performs better in attack detection than the $\chi^2$ test, i.e., DeeBBAA attacks can more comfortably hide when the $L_2$ norm of the residues are considered for hypothesis testing. It loosely implies that on average the DeeBBAA attacks inject deviations in both positive and negative directions, intelligently, leading to an overall balanced residue vector, the squared sum of whose individual elements remain close to the pre-attack conditions.
\end{enumerate}
\subsubsection{Bypassing Advanced Defensive Systems}\label{subsec:res_adv_defense}

\begin{figure*}[!ht]
    \centering
    \includegraphics[keepaspectratio, width=0.8\textwidth]{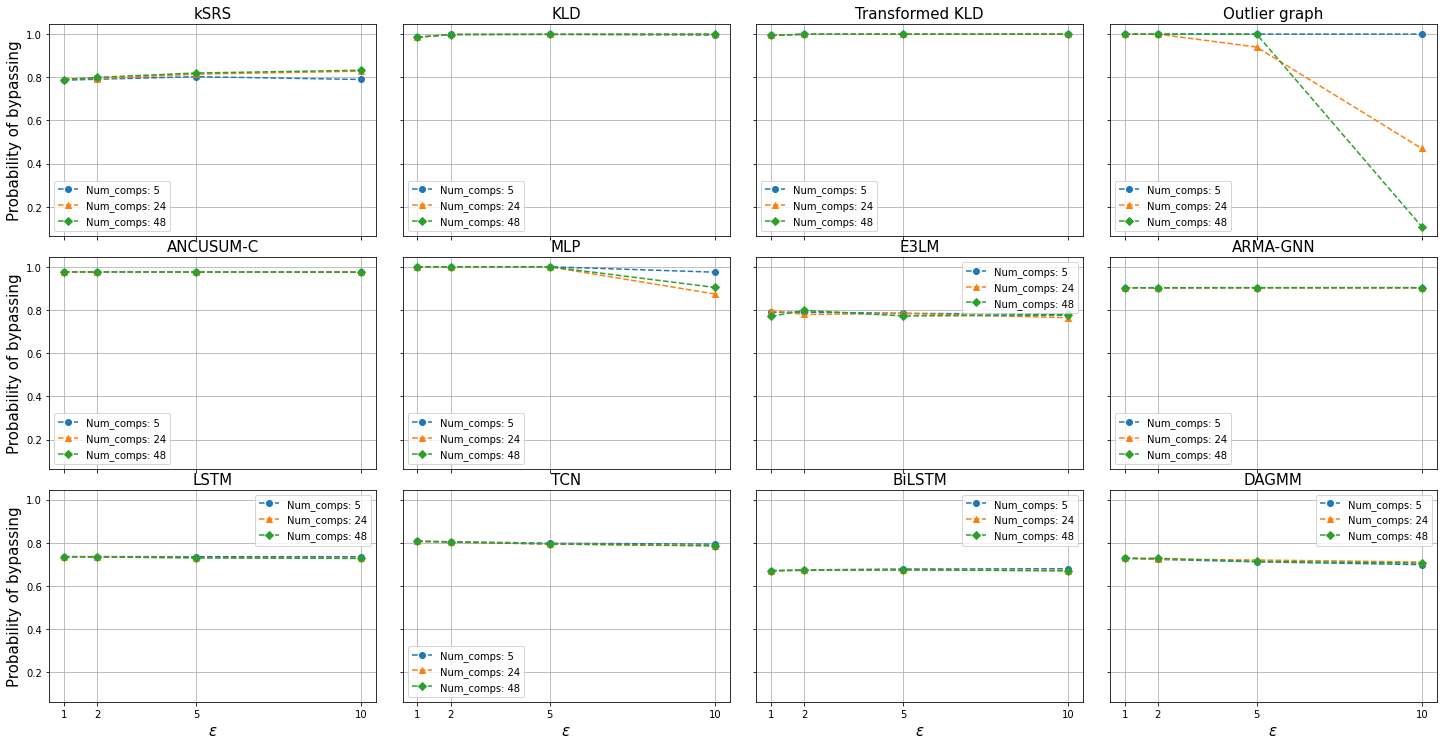}
    \caption{Probability of DeeBBAA attacks bypassing statistical consistency and learning based defenses for IEEE 39 bus  system under localized attack region with different values of epsilons and number of compromised measurements.}
    \label{fig:39loc_defense_comp}
\end{figure*}

\begin{figure*}[!ht]
    \centering
    \includegraphics[keepaspectratio, width=0.8\textwidth]{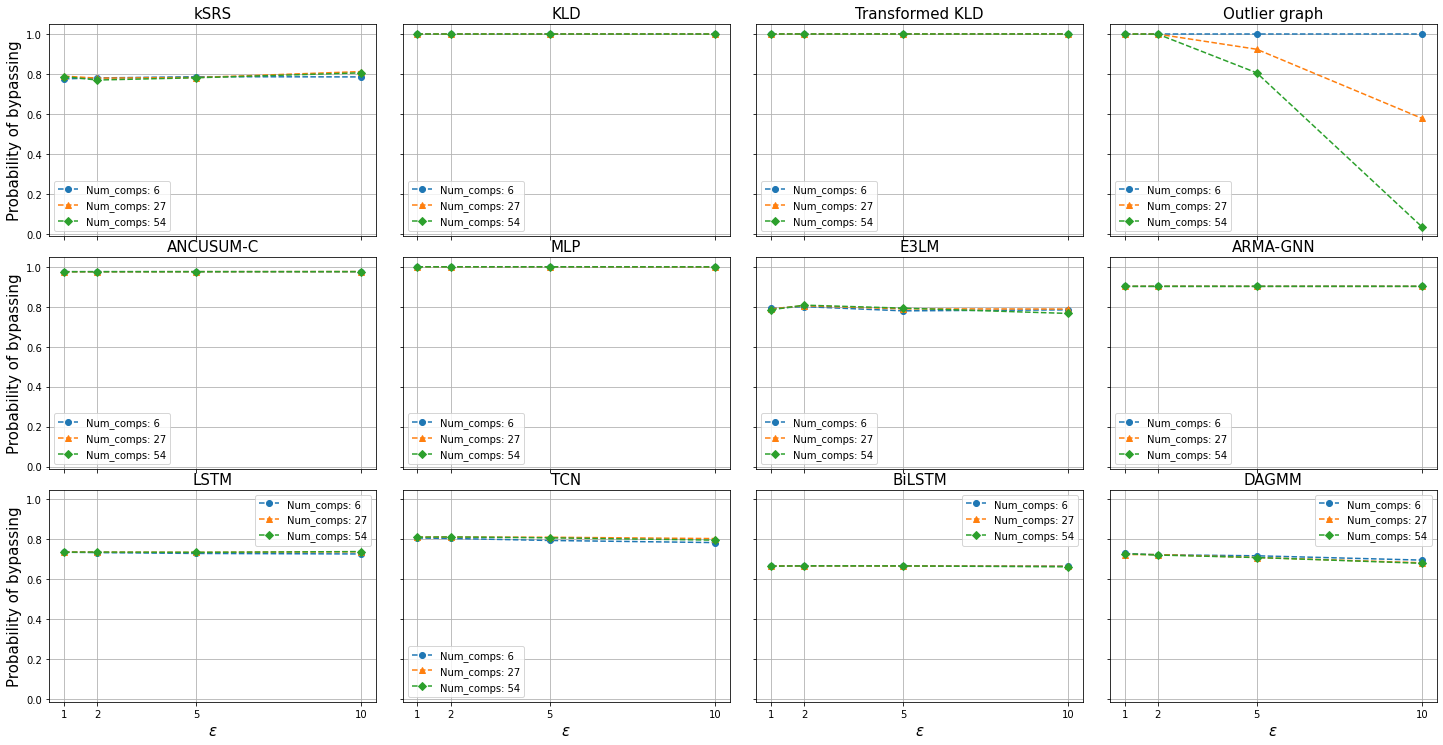}
    \caption{Probability of DeeBBAA attacks bypassing statistical consistency and learning based defenses for IEEE 39 bus  system under delocalized attack region with different values of epsilons and number of compromised measurements.}
    \label{fig:39deloc_defense_comp}
\end{figure*}

\begin{figure*}[!ht]
    \centering
    \includegraphics[keepaspectratio, width=0.8\textwidth]{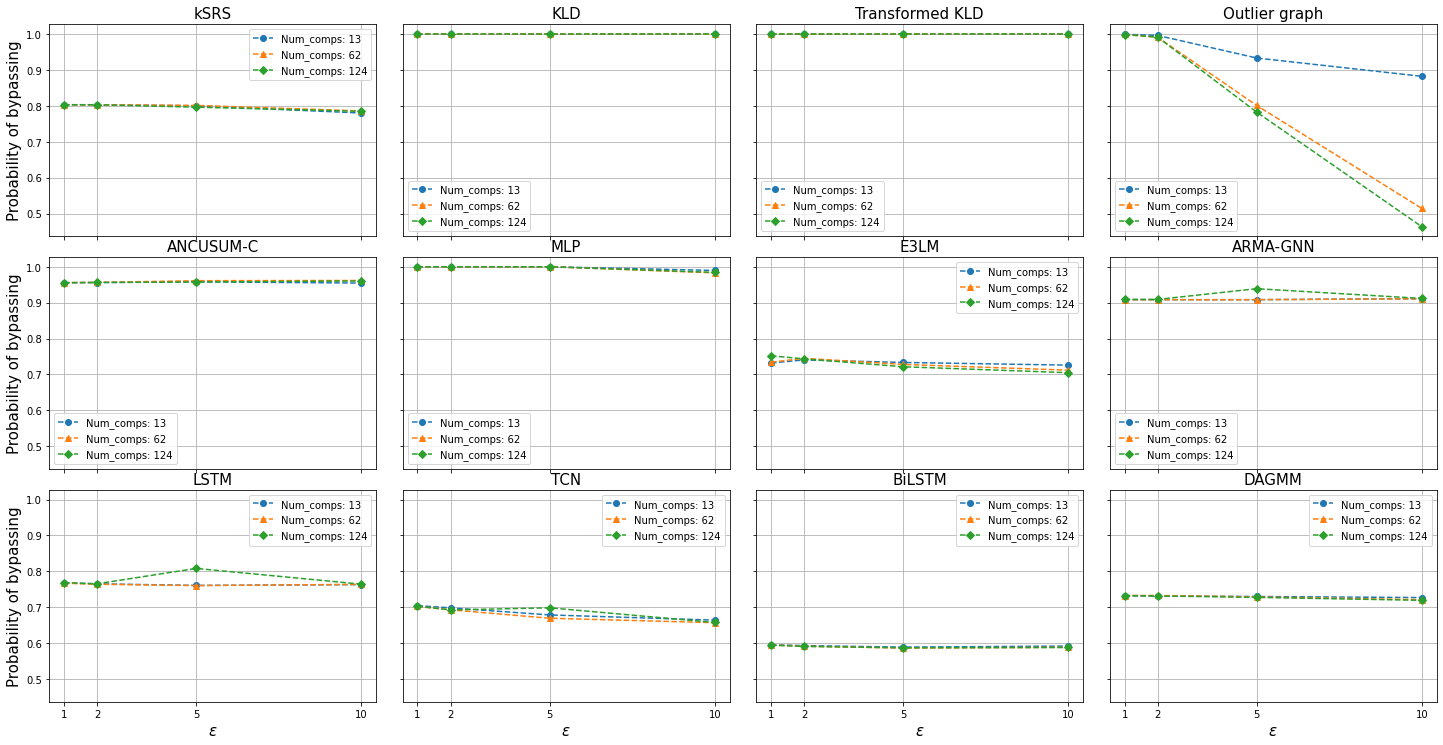}
    \caption{Probability of DeeBBAA attacks bypassing statistical consistency and learning based defenses for IEEE 118 bus  system under localized attack region with different values of epsilons and number of compromised measurements.}
    \label{fig:118loc_defense_comp}
\end{figure*}

\begin{figure*}[!ht]
    \centering
    \includegraphics[keepaspectratio, width=0.8\textwidth]{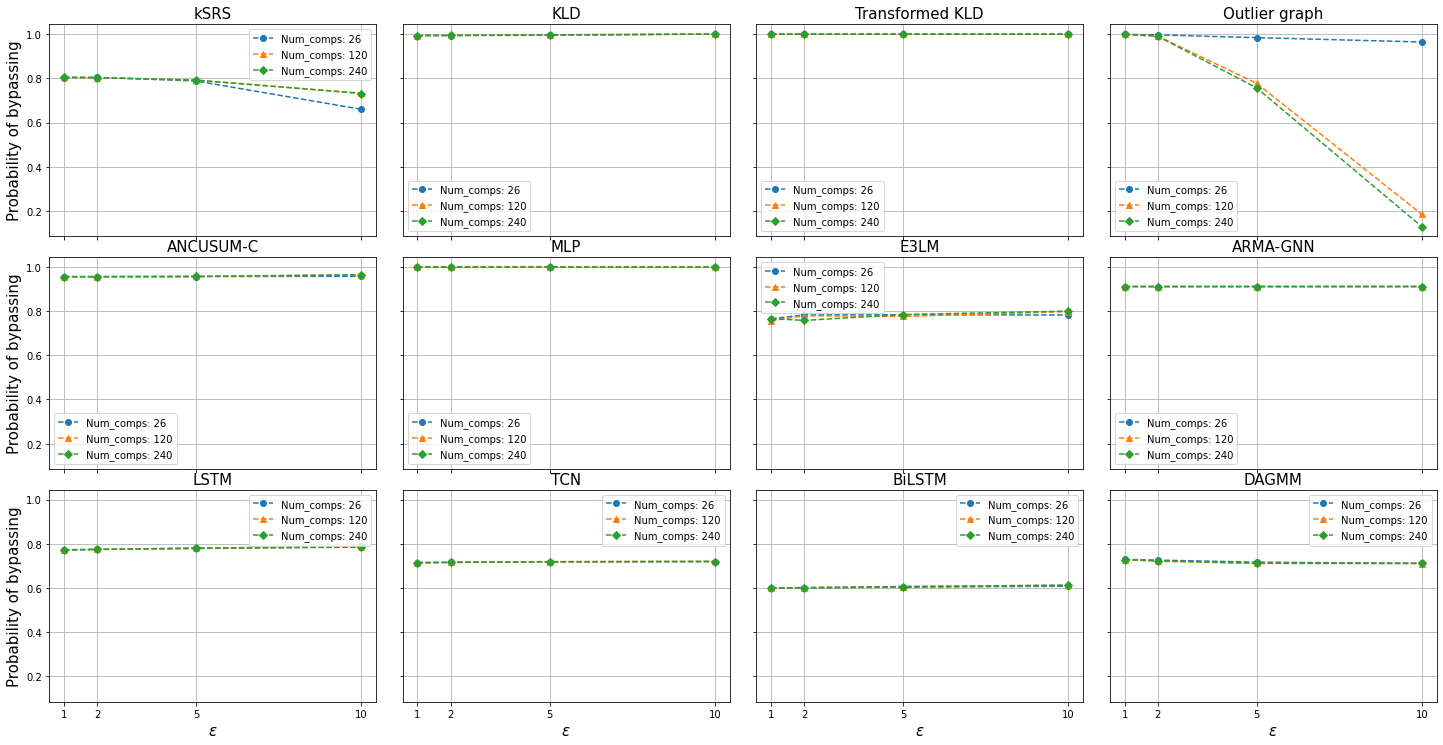}
    \caption{Probability of DeeBBAA attacks bypassing statistical consistency and learning based defenses for IEEE 118 bus  system under delocalized attack region with different values of epsilons and number of compromised measurements.}
    \label{fig:118deloc_defense_comp}
\end{figure*}

\begin{figure*}[!ht]
    \centering
    \includegraphics[keepaspectratio, width=\textwidth]{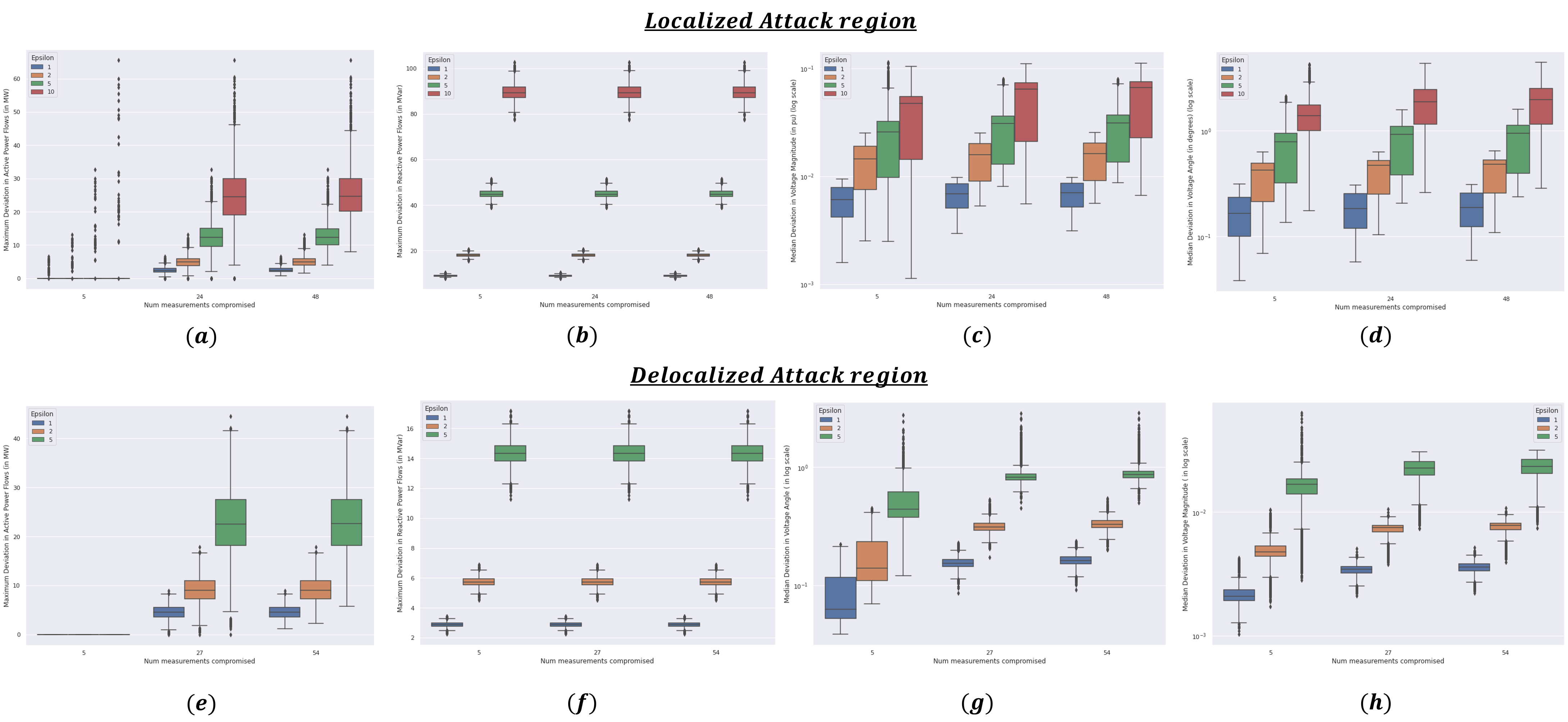}
    \caption{Deviation statistics for IEEE 39 bus test system. Number of measurements compromised by the attack vector is plotted on the x-axis and y-axis consists of the deviations. For each of the values in x-axis, there are 4 boxplots each corresponding to a value of $\epsilon$ as shown in the legends. Subplots a, b, c,d correspond to the localized attack region setting and plots e, f, g and h correspond to the delocalized attack region case. (a,e). Maximum Deviation in real power measurements, (b,f). Maximum deviation in reactive power measurements, (c,g). Median Deviation in Voltage magnitude estimates (in log scale), (d,h). Median deviation in voltage angle estimates (in log scale).}
    \label{fig:39bus_dev_stats}
\end{figure*}

\begin{figure*}[!ht]
    \centering
    \includegraphics[keepaspectratio, width=\textwidth]{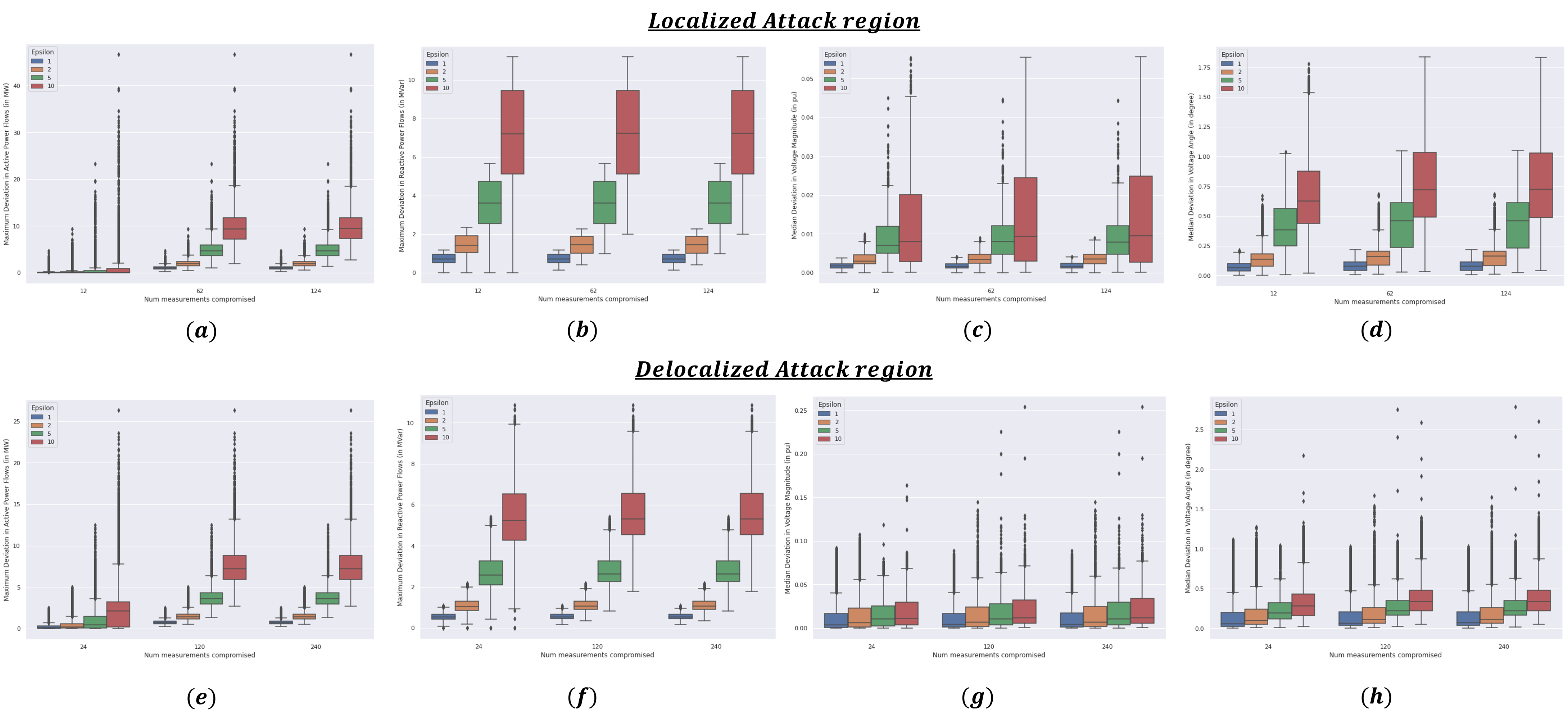}
    \caption{Deviation statistics for the IEEE 118 bus test system. Number of measurements compromised by the attack vector is plotted on the x-axis and y-axis consists of the deviations. For each of the values in x-axis, there are 4 boxplots each corresponding to a value of $\epsilon$ as shown in the legends. Subplots a, b, c,d correspond to the localized attack region setting and plots e, f, g and h correspond to the delocalized attack region case. (a,e). Maximum Deviation in real power measurements, (b,f). Maximum deviation in reactive power measurements, (c,g). Median Deviation in Voltage magnitude estimates (in log scale), (d,h). Median deviation in voltage angle estimates (in log scale).}
    \label{fig:118bus_dev_stats}
\end{figure*}

\begin{figure*}[!ht]
    \centering
    \includegraphics[keepaspectratio, width=\textwidth]{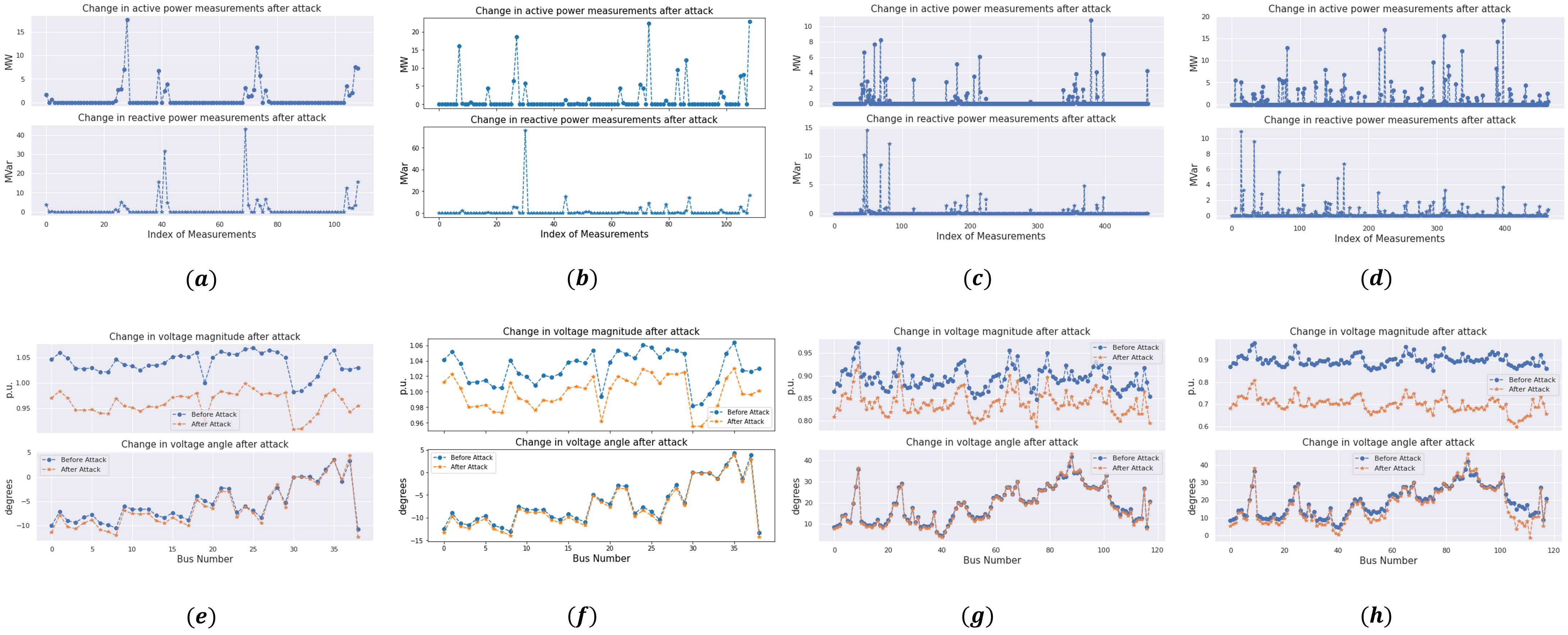}
    \caption{Point Deviation for the IEEE 39 and 118 bus test systems.}
    \label{fig:39_118 point_dev}
\end{figure*}
The probability of DeeBBAA attacks bypassing advanced defense mechanisms including physics-inspired statistical methods and learning based algorithms as outlined in Section \ref{subsec:exp_defense} are plotted in Figures \ref{fig:39loc_defense_comp} and \ref{fig:39deloc_defense_comp} for the IEEE 39 bus test system under localized and delocalized attack region settings respectively. Similar results for the IEEE 118 bus test system under localized and delocalized settings are plotted in Figures \ref{fig:118loc_defense_comp} and \ref{fig:118deloc_defense_comp} respectively. In each of these figures there are 12 subplots, each corresponding to a particular defense baseline specified in their respective titles. For each of these subplots, the probability of bypassing the concerned defense baseline is presented for different values of $\epsilon$ and the number of compromised measurements. The first 5 subplots correspond to the statistical methods of defense and the remaining 7 correspond to learning based detectors. Following are the observations and inferences drawn from these results:
\begin{enumerate}
    \item DeeBBAA attacks can bypass the KLD and Transformed KLD approaches with almost 100\% probability in all the considered scenarios. This shows that DeeBBAA attacks do not cause any significant deviations in the empirical distribution of the measurement vector. Additionally, the real-time quickest detection approaches like ANCUSUM-C also fails to detect DeeBBAA attacks for more than 95\% of instances in all the considered scenarios.
    \item Amongst the statistical methods considered, only the outlier graph based anomaly detection algorithm is able to detect DeeBBAA attacks of high magnitude and high spread which are generated when all the measurement units in an attack region are injected with the DeeBBAA attack vector formed with $\epsilon = 10$. Under all the remaining circumstances, i.e., when a subset of the measurements in an attack region are targeted or when the value of $\epsilon$ is less than 10, DeeBBAA successfully bypasses the outlier graph based detector with high probabilities of greater than 80\%. This is because as the spread and magnitude of DeeBBAA attacks simultaneously increase, the number of nodes and edges also increases in the outlier graph that leads to increased chances of detection.
    \item Amongst the learning based methods, the MLP completely fails to detect DeeBBAA under all circumstances. This is reasonable as the substitute NSE model is also an MLP but with a different architecture. However, DeeBBAA bypasses powerful time-series networks like LSTM, Bi-LSTM and TCN with average probability of 78\%, 70\% and 80\% respectively for the IEEE 39 bus system and with average probability of 76\%, 60\% and 70\% for the IEEE 118 bus system. The probability of bypassing the Bidirectional LSTM is the least in all cases but is still appreciable as Bi-LSTMs are extremely powerful networks that can detect anomalous patterns in a time series while no special effort was made in making the DeeBBAA attacks consistent in time.
    \item The unsupervised energy based DAGMM algorithm is evaded by the DeeBBAA attacks with probability of 70\% or more in all the cases.
    \item In the localization front, E3LM performs the best in terms of defense. For the IEEE 39 bus test system, DeeBBAA bypasses E3LM around 80\% of the time in all cases as compared to bypassing ARMA-GNN 90\% of the times. Similarly for the 118 bus test system, DeeBBAA bypasses E3LM with probability of 70\% or more in the localized attack setting and of around 80\% in the delocalized attack setting. Whereas in both the cases the probability of DeeBBAA evading ARMA-GNN is around 90\% on average. It is an appreciable result noting the fact that E3LM consists of an ensemble of 200 extreme learning machines that makes it an extremely powerful pattern detector. 
    \item The most interesting fact is that the NSE mimicking the AC-PSSE is a simple Multi-Layer Perceptron and is trained using a completely different dataset, Dataset A, than what is used to train the defense baselines. Moreover, the online DeeBBAA attacks are formulated using the same dataset, Dataset B that is used to train the defense baselines, while using no information about the defense baselines. Note that this represents the worst case scenario from the attacker's perspective where the defenses that it has to bypass are trained to identify generic SFDIA using a dataset (Dataset B) on which it has to implement online attacks without any sort of re-training while DeeBBAA is trained offline on a completely different dataset (Dataset A) with no overlaps. Under these circumstances DeeBBAA is able to bypass the defense baselines with probabilities ranging between 70\%-100\% in most of the cases.
\end{enumerate}

\subsubsection{Deviations in state estimates and measurements}

Recall that since the input to the NSE are just power measurements, i.e. bus injections and line flows, the DeeBBAA attacks hence generated are injected only into power measurements. In this part, the deviations caused by the DeeBBAA injections in power measurements and the corresponding deviations caused in the states estimated from the manipulated measurements are analysed. Figures \ref{fig:39bus_dev_stats} and \ref{fig:118bus_dev_stats} consist of boxplots depicting the distribution of the maximum deviation caused in the active and reactive power measurements and the median deviations caused in the state estimates after DeeBBAA attacks for the IEEE 39 bus and 118 bus cases respectively. The first row of each of these figures correspond to the localized attack region setting and the second row to the delocalized attack region setting. The x-axis of each subplot consists of the number of measurements compromised and groups of four boxplots corresponding to each value in the x-axis represent the four values of $\epsilon$. Deviations in estimated voltage magnitude and angles for the IEEE 39 bus case in Figures \ref{fig:39bus_dev_stats}.c, \ref{fig:39bus_dev_stats}.d, \ref{fig:39bus_dev_stats}.g, \ref{fig:39bus_dev_stats}.h are plotted on a log scale for better visualization. Fig. \ref{fig:39bus_dev_stats}.a and b represents the maximum deviation in real and reactive power measurements caused by DeeBBAA. Fig \ref{fig:39bus_dev_stats}.c and d represent the median deviations in estimated voltage magnitude and angles in the post attack scenario. Corresponding images in the second row of Fig.\ref{fig:39bus_dev_stats} present identical results for the delocalized scenario. The exact same structure is followed in Fig. \ref{fig:118bus_dev_stats}. As the value of epsilon increased the range of deviations and their median value also increases. In the localized attack region setting, DeeBBAA can introduce as high as 0.1 per unit median deviations in the estimated voltage magnitude and more than 2 degree median deviations in the estimated angles for both the IEEE 39 and 118 bus cases. In the delocalized attack region setting the deviations induced in states are higher with more than 1 per unit median deviations in the estimated voltage magnitude for the IEEE 39 bus system and more than 0.2 per units for the IEEE 118 bus system. Figure \ref{fig:39_118 point_dev} shows the actual changes in active and reactive power measurements and estimated states for one particular instance for the IEEE 39 and 118 bus test systems before and after the DeeBBAA attacks. The four subplots on the top row represent the changes in active and reactive power measurements before and after attack for the IEEE 39 bus case in the localized setting, IEEE 39 bus case in the delocalized setting, IEEE 118 bus case in the localized setting and IEEE 118 bus case in the delocalized setting. The x-axis in each of these subplots represent the index set of all the available measurements in their respective networks. As can be observed, the number of measurement units having positive deviation is a small subset of the total number of measurements available in the network. In the second row subplots, a before and after attack representation of the estimated states are presented for the IEEE 39 bus case in the localized setting, IEEE 39 bus case in the delocalized setting, IEEE 118 bus case in the localized setting and IEEE 118 bus case in the delocalized setting. The x-axis in each of these plots represent the index set of bus numbers, $\mathcal{N}$. In both the IEEE 39 bus and 118 bus cases, the deviations incurred by the attacks in the state estimates are greater in the delocalized attack region setting as compared to the localized one.

\section{Conclusion}\label{sec:conclusion}
Stealthy False Data Injection Attacks pose a major threat to the operational integrity of vital cyber-physical networks, the smart power system being one of the most crucial examples owing to its contribution to global economic development. While newer SFDIA design approaches focus only on bypassing the conventional residue-based BDD safeguards, even in the wake of newer and more sophisticated algorithms for SFDIA detection, it becomes necessary to rigorously ascertain the effectiveness and highlight the vulnerabilities and limitations of these state-of-the-art defensive systems in order to prevent false complacency. In the wake of these circumstances, this work proposes a benchmark SFDI attack framework, called DeeBBAA that takes a fundamentally different approach than the existing attack design mechanisms in order to facilitate the development of stronger defense paradigms to ensure the security of the power systems against intelligent adversaries. 

We show how by leveraging only historical data pertaining to a small subset of components from an unknown power system, an adversary can launch strong, versatile, and highly evasive FDI attacks that can lead to significant deviations in the state estimates of the network using DeeBBAA, all while requiring no information on the underlying power system, the target state estimation operation or the defensive algorithms put in place by the power system operator. Using a fast convex relaxation of the adversarial optimization problem against a substitute regression model, partially mimicking the operation of an unknown power system state estimator, DeeBBAA adversarially perturbs the real-time power measurements in a stealthy manner that allows them to evade detection by not only conventional BDD but also a variety of state of the art statistical and data-driven attack detection mechanisms. The adversary can tune the strength of the attack either directly by using a scaling factor or by changing the size of the attack region that it is targeting. Flexible attack regions are considered to further simplify reconnaissance for the adversary. 

The analysis presented in this work uncovers a grave situation that highlights the extreme vulnerability of attack detection mechanisms in the face of a breach in data or communication channel security. It immediately follows that in order to ensure the sustainability of security measures in cyber-physical systems, a combination of means for data security and communication channel security needs to be studied along with stronger attack detection mechanisms, and even the slightest of slacks in any one of these can lead to harmful consequences for the underlying network. 

\section{Supplementary Material}
\subsection{Details of Attack Regions for the IEEE 39 and 118 bus test systems}
\begin{figure}[htp]
\centering
\begin{subfigure}{\columnwidth}
\centering
\includegraphics[keepaspectratio,width=0.85\columnwidth]{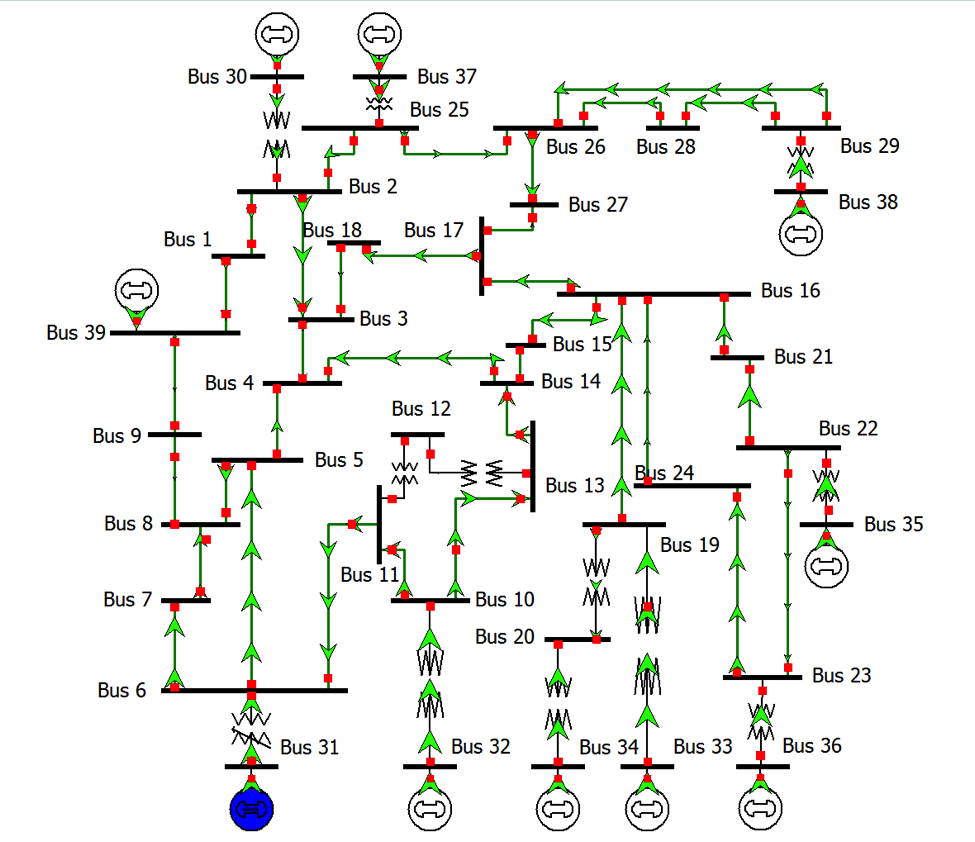}
\caption{IEEE 39 bus test system.}
\label{fig:ieee39}
\end{subfigure}

\begin{subfigure}{\columnwidth}
\centering
\includegraphics[keepaspectratio,width=0.85\columnwidth]{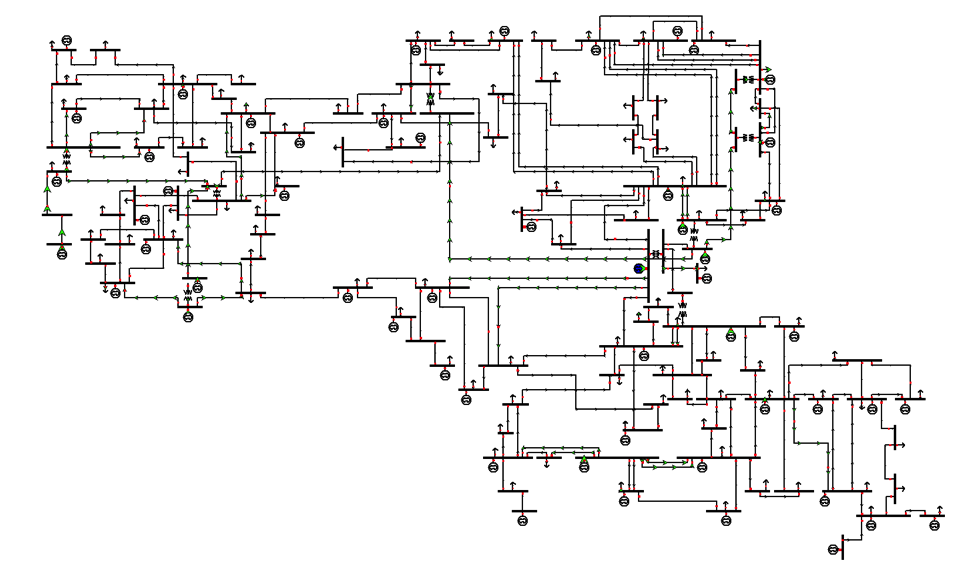}
\caption{IEEE 118 bus test system.}
\label{fig:ieee118}
\end{subfigure}
\caption{IEEE standard test systems used for experiments.}

\label{fig:res_dist}
\end{figure}

The IEEE 39 bus test system\cite{4113518}, depicted in Fig. \ref{fig:ieee39} is characterized by 39 nodes, 10 generators and 46 branches. The IEEE 118 bus test system, depicted in Fig. \ref{fig:ieee118}, represents a simple approximation of the American Electric Power system (in the U.S. Midwest) as of December 1962 and consists of 19 generators, 35 synchronous condensers, 177 lines, 9 transformers, and 91 loads \cite{118bus}.  

The localized and delocalized attack regions considered for the IEEE 39 and 118 bus cases in the proposed work are elaborated in Table \ref{tab:att_reg_details}. The nodes are represented by their indices and the branches are represented by the tuple of nodes on which they are incident. The naming convention of the buses in the test systems follow the PandaPower conventions \cite{pandapower.2018}.

\begin{table*}[!ht]
\begin{center}
\caption{ Details of Attack Regions for the IEEE 39 and 118 bus test systems
}
\label{tab:att_reg_details}
\resizebox{\textwidth}{!}{
\renewcommand{\arraystretch}{1}
\begin{tabular}{|l|ll|ll|}
\hline
\textbf{IEEE Test System} & \multicolumn{2}{c|}{\textbf{Localized Attack Region}}                                                                                                                                                                                                                                                                                                                                                                     & \multicolumn{2}{c|}{\cellcolor[HTML]{FFFFFF}\textbf{Delocalized Attack Region}}                                                                                                                                                                                                                                                                                                                                                                                                                                                                                                                                                                                  \\ \hline
                          & \multicolumn{1}{c|}{$\mathcal{B}_A$}                                                                                                                                                          & \multicolumn{1}{c|}{$\mathcal{E}_A$}                                                                                                                                                                                      & \multicolumn{1}{c|}{$\mathcal{B}_A$}                                                                                                                                                                         & \multicolumn{1}{c|}{$\mathcal{E}_A$}                                                                                                                                                                                                                                                                                                                                                                                                              \\ \hline
\textbf{39 bus}           & \multicolumn{1}{l|}{\begin{tabular}[c]{@{}l@{}}0, 1, 2, 24, \\ 25, 26, 27, 28\end{tabular}}                                                                                                   & \begin{tabular}[c]{@{}l@{}}(24, 25), (1, 24), \\ (0, 1), (1, 2), \\ (25, 26), (25, 27), \\ (25, 28), (27, 28)\end{tabular}                                                                                                & \multicolumn{1}{l|}{\begin{tabular}[c]{@{}l@{}}7, 8, 11, 17, \\ 26, 27, 30\end{tabular}}                                                                                                                     & \begin{tabular}[c]{@{}l@{}}(4, 7), (6, 7), (7, 8), \\ (25, 27), (27, 28), \\ (16, 26), (25, 26), \\ (2, 17), (16, 17), \\ (7, 8)\end{tabular}                                                                                                                                                                                                                                                                                                     \\ \hline
\textbf{118 bus}          & \multicolumn{1}{l|}{\begin{tabular}[c]{@{}l@{}}48, 36, 37, 38, \\ 40, 42, 43, 44, \\ 46, 47, 49, 50, \\ 51, 52, 56, 57, \\ 59, 62, 63, 66, \\ 67, 68, 70, 74,\\ 77, 80, 81, 117\end{tabular}} & \begin{tabular}[c]{@{}l@{}}(36 38), (42 43), \\ (43 44), (44 48), \\ (46 48), (46 68), \\ (47 48), (48, 49), \\ (48 50), (48 68), \\ (49 56), (50 51), \\ (50 57), (51 52), \\ (62 63), (68 74), \\ (74 117)\end{tabular} & \multicolumn{1}{l|}{\begin{tabular}[c]{@{}l@{}}5, 14, 15, 17, \\ 18, 21, 31, 32, \\ 33, 34, 38, 43,\\  44, 46, 49, 51, \\ 52, 69, 74, 75, \\ 79, 81, 83, 90, \\ 97, 100, 104, \\ 106, 107, 117\end{tabular}} & \begin{tabular}[c]{@{}l@{}}(20, 21), (46, 68), \\ (81, 82), (34, 36), \\ (43, 44), (31, 113), \\ (15, 16), (78, 79), \\ (79, 96), (79, 97), \\ (50, 51), (51, 52),\\  (75, 117), (69, 70),\\  (69, 74), (42, 43),\\  (18, 19), (68, 74), \\ (74, 117), (104, 107),\\  (49, 56), (33, 42), \\ (4, 5), (5, 6), (32, 36), \\ (16, 17), (107, 108), \\ (36, 38), (51, 52), \\ (100, 101), (105, 106), \\ (82, 83), (12, 14), \\ (13, 14)\end{tabular} \\ \hline
\end{tabular}
}\end{center}
\end{table*}

\subsection{EigenValues of W}

\begin{figure}[!ht]
    \centering
    \includegraphics[ keepaspectratio,width=\columnwidth]{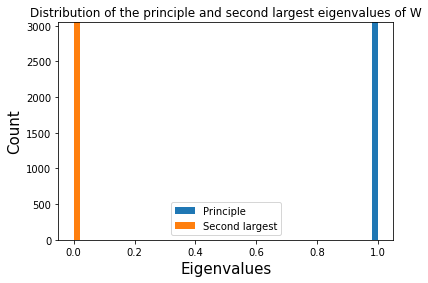}
    \caption{Distribution of eigenvalues of $\pmb{W}$.}
    \label{fig:eigval_dist}
\end{figure}

As stated in Section 3.3 of the main manuscript, here we show that the principal eigenvalue of $\pmb{W}$ remains close to one and the second largest value is very close to 0. We plot a bar graph of the first and second largest eigenvalues of $\pmb{W}$ computed over 3000 datapoints and the result is shown in  Fig. \ref{fig:eigval_dist}. As can be clearly seen, the principle eigenvalues are all very close to 1 and the second largest eigenvalues are infinitesimally close to zero implying that $\pmb{W}$ is an approximately rank one matrix.

\subsection{Training the Defensive Baselines}

This section is split into three parts each corresponding to one of the three types of defense algorithms used in this work:

\begin{figure}[htp]
\centering
\begin{subfigure}{\columnwidth}
\centering
\includegraphics[keepaspectratio,width=0.85\columnwidth]{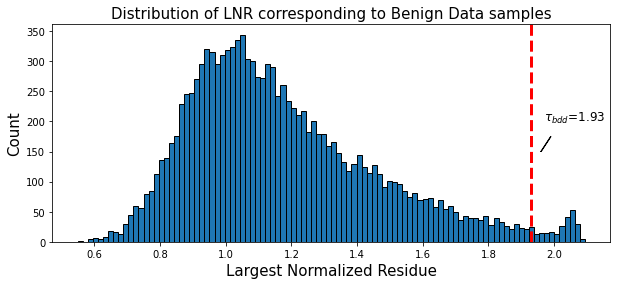}
\caption{Distribution of Largest Normalized Residues or the $L_{\infty}$ norm of residue vectors for benign data samples in the IEEE 39 bus test system.}
\label{fig:res_39}
\end{subfigure}

\begin{subfigure}{\columnwidth}
\centering
\includegraphics[keepaspectratio,width=0.85\columnwidth]{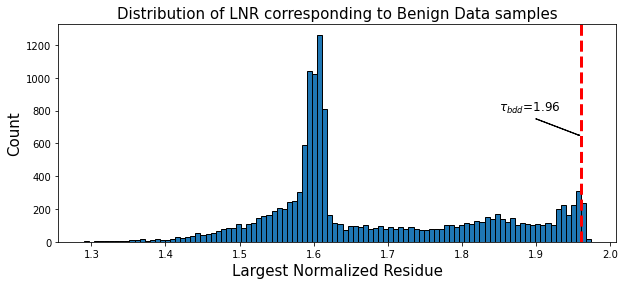}
\caption{Distribution of Largest Normalized Residues or the $L_{\infty}$ norm of residue vectors for benign data samples in the IEEE 118 bus test system.}
\label{fig:res_118}
\end{subfigure}
\caption{Distribution of Largest Normalized residues corresponding to benign data samples}

\label{fig:res_dist}
\end{figure}

    \begin{figure}[htp]
\centering
\begin{subfigure}{\columnwidth}
\centering
\includegraphics[keepaspectratio,width=0.85\columnwidth]{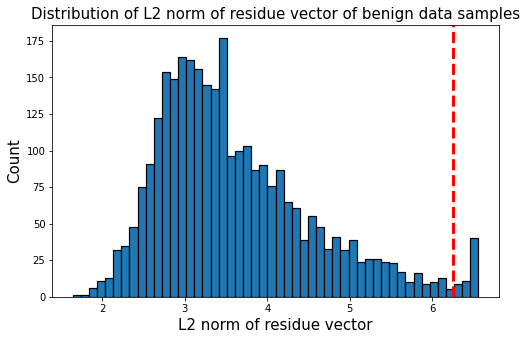}
\caption{Distribution of $L_{2}$ norm of residue vectors for benign data samples in the IEEE 39 bus test system.}
\label{fig:res_39_L2}
\end{subfigure}

\begin{subfigure}{\columnwidth}
\centering
\includegraphics[keepaspectratio,width=0.85\columnwidth]{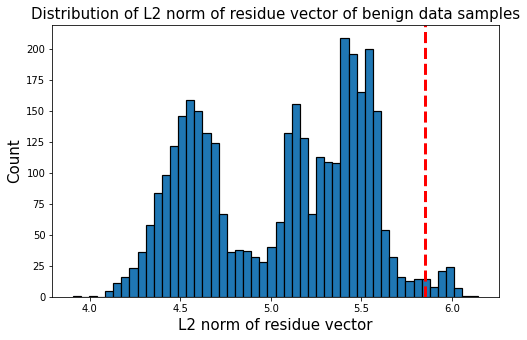}
\caption{Distribution of $L_{2}$ norm of residue vectors for benign data samples in the IEEE 118 bus test system.}
\label{fig:res_118_L2}
\end{subfigure}
\caption{Distribution of L2 norm of residues corresponding to benign data samples}

\label{fig:res_dist}
\end{figure}

\begin{itemize}
    \item \textit{Conventional BDD}: As part of conventional BDD algorithms, LNRT and $\chi^2$ tests were carried out. Training of these algorithms involve identifying the BDD thresholds or $\tau_{\infty}$ and $\tau_{2}$ respectively as given in Equations 8 and 9 in the main manuscript. Dataset B consisting of good data representing normal operating conditions is used for this purpose. For each of the data samples in Dataset B, the L$\infty$ norm and L2 norm of the normalized residue vector is computed using Eq. 7 of the main manuscript. Keeping false alarm rates at $2\%$, the L$\infty$ norm and L2 norm obtained as the 98 percentile values are used as $\tau_{\infty}$ and $\tau_{2}$ respectively. Figures \ref{fig:res_39} and \ref{fig:res_118} depict the distribution of the $L_\infty$ norm of the normalized residues corresponding to benign data points for the IEEE 39 bus and 118 bus test systems respectively. Figures \ref{fig:res_39_L2} and \ref{fig:res_118_L2} depict the distribution of the L$2$ norm of the normalized residues corresponding to benign data points for the IEEE 39 bus and 118 bus test systems respectively. The threshold values are marked as vertical lines in the plots.
    
    \item \textit{Statistical Methods}: Five state of the art statistical consistency tests, namely the \textbf{KLD} test\cite{7035067}, jointly\textbf{Transformed KLD} test\cite{7961272}, \textbf{kSRS} test\cite{9676996}, \textbf{ANCUSUM-C}\cite{9311640} and \textbf{Outlier Graph} based test\cite{9107408}  are considered as defense baselines. Training, with respect to the KLD test, Transformed KLD test, kSRS test  and ANCUSUM-C test, imply computation of thresholds to be used for hypothesis testing using data under normal operating conditions. Following the procedures elaborated in the respective papers, Dataset B is used for threshold computations of the aforementioned algorithms. For the Outlier Graph based SFDIA detection algorithm, thresholds are computed using Dataset $B_{att}$ following the procedure outlined in \cite{9107408}.
    
    \item \textit{Learning Based Methods}: Six supervised and one unsupervised state of the art learning based SFDIA detection algorithms are considered. A \textbf{MLP} classifier is used to solve a binary classification problem whose output indicates whether the power system is compromised or not. The MLP has three hidden layers of size 256, 128 and 64 respectively followed by a single unit output layer with sigmoid activation. The hidden layer outputs are followed by a dropout regularization with drop probability of 10\% and a Leaky-ReLU activation function. The input to the network is the measurement vector, consisting of all power system measurements including the voltage magnitudes at each bus.
    
    For the sequence to sequence time series networks, i.e., \textbf{LSTM} \cite{9773032, 8334607, 9542963}, \textbf{Bi-LSTM} and \textbf{TCN}\cite{9557319, 9049087, 8233155}, inputs are a sequence of measurement vectors over a time window of length 50 units. The output are binary labels of the same sequence length each representing whether the input measurements at a particular timestep is compromised or not. The LSTM and Bi-LSTM networks have 3 hidden layers of dimensionality 64 each with default activations, while the output layer has a sigmoid activation. The architecture of the TCN is inspired from \cite{tcn-arch}. Number of filters used were 256 with a kernel size of 5 and number of stacks fixed at 3. Dilations of values 1, 2 and 4 were used. For the localization task, \textbf{E3LM}\cite{9055170} and \textbf{ARMA-GNN}\cite{9559412} were considered as baselines. The hyperparameters of these algorithms were kept the same as in the respective works that proposed them. For the localization task, the target variable is an array of 0s or 1s depending on which state variables are compromised. For every attack data sample in Dataset $B_{att}$, the target variable in these cases is a vector with dimensions equal to that of the state perturbation $\pmb{c}$. If the $k$-th element of $\pmb{c}$ is non-zero, then the $k$-th element of the target variable is 1 otherwise 0. A train-test split of 80-20 is carried out on Dataset $B_{att}$ to train these algorithms. They are trained till the  accuracy of classification of all these algorithms on the test sets is >98\%. Finally,  \textbf{DAGMM}\cite{9706368} is implemented following https://github.com/tnakae/DAGMM repository and keeping the same hyperparameters given in the paper. 
\end{itemize}

\bibliographystyle{ieeetr}
\bibliography{main.bib}

\end{document}